\documentclass[12pt]{article}

\usepackage[margin=1in]{geometry}

\usepackage{enumitem} 

\usepackage{times}

\usepackage{setspace} 
\usepackage{caption}
\renewcommand{\arraystretch}{0.8} % Strech matrices and tables back to regular spacing

\usepackage{amsmath}
\usepackage{amssymb}
\usepackage{mathtools} % matrix*: [c] centered, [l] left-aligned, [r] right-aligned

\usepackage[utf8]{inputenc} 
\usepackage[english]{babel}

\usepackage{graphicx} 

\usepackage{float}

\usepackage{siunitx} 
\newtheorem{theorem}{Theorem}
\newtheorem{assumption}{Assumption}

\usepackage[title]{appendix}

\numberwithin{equation}{section}

\newcommand\blfootnote[1]{
  \begingroup
  \renewcommand\thefootnote{}\footnote{#1}
  \addtocounter{footnote}{-1}
  \endgroup
}

\usepackage[hang]{footmisc}
\usepackage{hyperref}
\hypersetup{pdftex,colorlinks=true, allcolors=black}

\usepackage[round,sort,comma]{natbib} 
\begin{document}

\begin{titlepage}
\begin{center}
%\linespread{1.2}
\vspace*{1cm}
\Large{\textbf{A mixture autoregressive model based on Gaussian and Student's \textit{t}-distributions}} \\

\vspace{1cm} 
\Large{Savi Virolainen}\\
\large{University of Helsinki}\\
\vspace{2cm}

\begin{abstract}
\noindent We introduce a new mixture autoregressive model which combines Gaussian and Student’s $t$ mixture components. The model has very attractive properties analogous to the Gaussian and Student’s $t$ mixture autoregressive models, but it is more flexible as it enables to model series which consist of both conditionally homoscedastic Gaussian regimes and conditionally heteroscedastic Student’s $t$ regimes. The usefulness of our model is demonstrated in an empirical application to the monthly U.S. interest rate spread between the 3-month Treasury bill rate and the effective federal funds rate.\\[1cm]

%Gaussian distribution, Student's $t$-distribution
\end{abstract}

\vfill

\blfootnote{This work was supported by the Academy of Finland under Grant 308628.}
\blfootnote{Contact address: Savi Virolainen, Faculty of Social Sciences, University of Helsinki, P. O. Box 17, FI–00014 University of Helsinki, Finland; e-mail: savi.virolainen@helsinki.fi.}
\blfootnote{The author has no conflict of interest to declare.}

\end{center}
\end{titlepage}

\section{Introduction}
Recently, \citet*{Kalliovirta+Meitz+Saikkonen:2015} introduced a mixture autoregressive model based on Gaussian distribution with very attractive features. The Gaussian mixture autoregressive (GMAR) model has linear Gaussian autoregressions as its component models and mixing weights that, for a $p$th order model, depend on the full distribution of the $p$ past observations. The specific formulation of the mixing weights leads to ergodicity and full knowledge of the stationary distribution of $p+1$ consecutive observations. Moreover, it allows regime switches to depend on the level, variability, and temporal dependence of the past observations.

\cite*{Meitz+Preve+Saikkonen:2018} proposed a mixture autoregressive model closely related to the GMAR model but based on Student's $t$-distribution. The Student's $t$ mixture autoregressive (StMAR) model has linear Student's $t$ autoregressions as its component models and mixing weights constructed analogously to the GMAR model, leading to similar theoretical and practical properties. The linear Student's $t$ autoregressions have the same form for the conditional mean as the Gaussian autoregressions (a linear function of the past observations) but different conditional variance. In particular, the conditional variances of the Student’s $t$ autoregressions depend on quadratic forms of past observations, whereas in the Gaussian case the conditional variances of the component models are constants. Utilization of the $t$-distribution does hence not only allow the StMAR model to account for larger kurtosis than the GMAR model but also stronger forms of conditional heteroskedasticity. 

In this paper, we propose a generalization of the GMAR and StMAR models. The G-StMAR model accommodates both Gaussian autoregressions and Student's $t$ autoregressions as its component models, and its mixing weights are constructed analogously to the GMAR and StMAR models, leading to similar attractive features. It thus enables to model series which consist of regimes with time varying conditional variance and excess kurtosis as well as regimes with constant conditional variance and zero excess kurtosis. It turns out that the G-StMAR model is a limiting case of a StMAR model with the $t$-distributions of some regimes tending to normal distributions as the degrees of freedom parameters tend to infinity. As opposed to the limiting StMAR model, the advantage of the G-StMAR model is that it removes the redundant degrees of freedom parameters from the model and is free from numerical problems induced by weak identification of very large degrees of freedom parameters.

We demonstrate the usefulness of the G-StMAR model in an empirical application to the monthly U.S. interest rate spread between the 3-month Treasury bill (TB) rate and the effective federal funds (FF) rate. Our G-StMAR model identifies three regimes for the spread, with a GMAR type regime mainly appearing after the financial crisis in 2008 when the zero lower bound limits movements of the spread. The remaining regimes are of the StMAR type, one accommodating eras of low mean and high variability and the other high mean and moderate variability. The former StMAR type regime dominates often when the market possibly anticipates decreases in the FF rate or has increased preference for safety, whereas the latter one mostly prevails when the Fed is arguably not expected to significantly decrease the FF rate target. Our findings are consistent with \citet*{Sarno+Thornton:2003} who found that the FF rate seems to adjust to the TB rate, supporting the hypothesis that the market anticipates movements of the FF rate, moving the TB rate, and hence the spread, in advance.

The rest of this article is organized as follows. Section \ref{sec:models} first introduces the component processes of the G-StMAR model and then proceeds to define the G-StMAR model and discusses its theoretical properties. Section \ref{sec:estim} discusses maximum likelihood (ML) estimation of the model parameters and establishes the asymptotic properties of the ML estimator. It is, in particular, discussed how the accompanying R package "uGMAR" \citep{uGMAR} estimates the model parameters in practice with a two-phase procedure. Section \ref{sec:building} describes a simple model selection procedure and discusses numerical consequences of very large degrees of freedom parameter estimates. Section \ref{sec:empexample} presents the empirical application to the interest rate spread and Section \ref{sec:conclusions} concludes. Details of the estimation procedure employed by uGMAR, as well as proofs for the stated theorems are given in an Appendix. 

Throughout this paper, we use the following notation. We write $\boldsymbol{x}=(x_1,...,x_n)$ for the column vector $\boldsymbol{x}$ where the components $x_i$ may be either scalars or (column) vectors. The notation $\boldsymbol{x}\sim n_d(\boldsymbol{\mu},\boldsymbol{\Gamma})$ signifies that the random vector $\boldsymbol{x}$ has a $d$-dimensional Gaussian distribution with mean $\boldsymbol{\mu}$ and (positive definite) covariance matrix $\boldsymbol{\Gamma}$. Similarly, $\boldsymbol{x}\sim t_d(\boldsymbol{\mu},\boldsymbol{\Gamma},\nu)$ signifies that $\boldsymbol{x}$ has a $d$-dimensional $t$-distribution with mean $\boldsymbol{\mu}$, (positive definite) covariance matrix $\boldsymbol{\Gamma}$, and degrees of freedom $\nu$ (assumed to satisfy $\nu>2$). The density functions and some properties of the multivariate Gaussian and Student's $t$-distributions are given in an Appendix. The vectorization operator $vec$ stacks columns of a matrix on top of each other and, $\iota_{d}$ is the $d$ dimensional vector $(1,0,...,0)$,  $I_d$ signifies the identity matrix of dimension $d$, and $\otimes$ denotes the Kronecker product. Moreover, $\boldsymbol{1}_d$ and $\boldsymbol{0}_d$ denote $d$ dimensional vectors of ones and zeros, respectively.

\section{Models}\label{sec:models}
We consider mixture autoregressive models in which each observation is generated by a mixture component that is randomly selected according to the probabilities pointed by the mixing weights. The mixture components are either (linear) conditionally homoscedastic Gaussian autoregressions as in the GMAR model \citep{Kalliovirta+Meitz+Saikkonen:2015} or conditionally heteroscedastic Student's $t$ autoregressions as in the StMAR model \citep{Meitz+Preve+Saikkonen:2018}. The mixing weights are functions of the past observations constructed in a way that, for a $p$th order model, leads to ergodicity and full knowledge of the stationary distribution $p+1$ consecutive observations. Moreover, as the mixing weights depend on the full distribution of the past $p$ observations, they allow regime switches to depend on the level, variability, and temporal dependence of the past observations. In this section, we first introduce the component processes of the G-StMAR model and then proceed define of the G-StMAR model and discuss its properties.

\subsection{Linear Gaussian and Student's \textit{t} autoregressions}\label{sec:gstmar_gausstud}
To develop theory and notation, we first consider the component processes of the G-StMAR model. For a linear $p$th order Gaussian or Student $t$ autoregression $z_t$, we have
\begin{equation}\label{eq:component}
z_t = \varphi_{0} + \sum_{i=1}^p\varphi_iz_{t-i} + \sigma_t\varepsilon_t, \quad \varepsilon_t \sim \text{IID}(0,1),
\end{equation} 
where $\sigma_t>0$, $\varphi_{0}\in\mathbb{R}$, and the autoregressive (AR) parameter $\boldsymbol{\varphi}=(\varphi_1,...,\varphi_p)$ satisfies the stationarity condition $\boldsymbol{\varphi}\in\mathbb{S}^p$ where
\begin{equation}\label{eq:sp}
\mathbb{S}^p = \lbrace (\varphi_1,...,\varphi_p)\in\mathbb{R}^p:1-\sum_{i=1}^p\varphi_{i}z^i\neq0 \text{ for } |z|\leq 1 \rbrace.
\end{equation}
In the case of Gaussian autoregression, the distribution of the errors terms $\varepsilon_t$ is standard normal and $\sigma_t$ is a constant $\sigma$ for all $t$. Denoting $\boldsymbol{z}_t=(z_t,...,z_{t-p+1})$ and $\mu=\text{E}[z_t]$, $\gamma_j=\text{Cov}(z_t,z_{t-j})$, and $\boldsymbol{\gamma}_p=(\gamma_1,...,\gamma_p)$, 
it is well know that the stationary solution to (\ref{eq:component}) for the Gaussian autoregression satisfies
\begin{align}
%\begin{aligned}
\boldsymbol{z}_t &\sim n_p(\mu\boldsymbol{1}_p,\boldsymbol{\Gamma}_p),\label{eq:gausdistr1}\\ 
(z_t,\boldsymbol{z}_{t-1}) &\sim n_{p+1}(\mu\boldsymbol{1}_{p+1},\boldsymbol{\Gamma}_{p+1}),\label{eq:gausdistr2}\\
z_t\mid\boldsymbol{z}_{t-1} &\sim 
n_1(\mu + \boldsymbol{\gamma}_p'\boldsymbol{\Gamma}_p^{-1}(\boldsymbol{z}_{t-1} - \mu\boldsymbol{1}_p),\gamma_0 - \boldsymbol{\gamma}_p'\boldsymbol{\Gamma}_p^{-1}\boldsymbol{\gamma}_p)=n_1(\varphi_0 + \boldsymbol{\varphi}'\boldsymbol{z}_{t-1},\sigma^2),\label{eq:gausdistr3}
%\end{aligned}
\end{align}
where $\mu=\varphi_0/(1-\boldsymbol{\varphi}'\boldsymbol{1}_p)$ $\boldsymbol{\gamma}_p=\boldsymbol{\Gamma}_p\boldsymbol{\varphi}$, and the covariance matrices $\boldsymbol{\Gamma}_{p}$ and $\boldsymbol{\Gamma}_{p+1}$ are Toeplitz matrices given as (see, e.g., L\"utkepohl (2005), eq. (2.1.39))
\begin{equation}
vec(\boldsymbol{\Gamma}_p) = (I_{p^2} - (\Phi\otimes\Phi))^{-1}\iota_{p^2}\sigma^2, \enspace
\Phi =
  \begin{bmatrix}
  \varphi_1\cdots\varphi_{p-1} &  \varphi_p \\
  I_{p-1} & \boldsymbol{0}_{p-1} \\
  \end{bmatrix},
\enspace
\boldsymbol{\Gamma}_{p+1} = 
   \begin{bmatrix}
  \gamma_0 & \boldsymbol{\gamma}_p' \\
  \boldsymbol{\gamma}_p & \boldsymbol{\Gamma}_p \\
  \end{bmatrix}.
\end{equation}

Using the same notation as in (\ref{eq:gausdistr1})-(\ref{eq:gausdistr3}) for $\boldsymbol{z}_{t-1}$, $\mu$, and $\boldsymbol{\Gamma}_p$, the Student's $t$ autoregressions utilized by \cite{Meitz+Preve+Saikkonen:2018} (which have also appeared at least in \cite{Spanos:1994} and \cite{Heracleus+Spanos:2006})
are obtained by letting $\varepsilon_t \sim t_1(0,1,\nu+p)$ with $\nu>2$ in (\ref{eq:component}) and defining 
\begin{equation}\label{eq:sigma_mt}
\sigma_t^2 = \frac{\nu - 2 + (\boldsymbol{z}_{t-1} - \mu\boldsymbol{1}_p)'\boldsymbol{\Gamma}_p^{-1}(\boldsymbol{z}_{t-1} - \mu\boldsymbol{1}_p)}{\nu-2+p}\sigma^2.
\end{equation}
This definition (which requires the stationarity condition of the AR parameter) guarantees stationarity of the Student's $t$ autoregressions.
Distributional properties of such stationary Student's $t$ autoregressions are similar to the Gaussian case, in particular \citep[Theorem 1]{Meitz+Preve+Saikkonen:2018},
\begin{align}
%\begin{aligned}
\boldsymbol{z}_t &\sim t_p(\mu\boldsymbol{1}_p,\boldsymbol{\Gamma}_p,\nu),\label{eq:studentdist1}\\
(z_t,\boldsymbol{z}_{t-1}) &\sim t_{p+1}(\mu\boldsymbol{1}_{p+1},\boldsymbol{\Gamma}_{p+1},\nu),\label{eq:studentdist2}\\
z_t\mid\boldsymbol{z}_{t-1} &\sim t_1(\varphi_0 + \boldsymbol{\varphi}'\boldsymbol{z}_{t-1},\sigma_t^2,\nu+p).\label{eq:studentdist3}
%\end{aligned}
\end{align} 
The aforementioned properties of the component processes are essential in the following discussions and will be exploited implicitly. Gaussian component processes of the G-StMAR model are referred to as \textit{GMAR type} and Student's $t$ component processes as \textit{StMAR type} since they are identical to the component processes of the GMAR model \citep{Kalliovirta+Meitz+Saikkonen:2015} and the StMAR model \citep{Meitz+Preve+Saikkonen:2018}, respectively.

\subsection{Gaussian and Student's \textit{t} mixture autoregressive model}\label{sec:gstmar}
Let $y_t$ ($t=1,2,...$) be the real valued time series of interest, and let $\mathcal{F}_{t-1}$ denote the $\sigma$-algebra generated by the random variables $\left\lbrace y_{t-j},j>0\right\rbrace$. For a G-StMAR model with $M$ mixture components and autoregressive order $p$, we have
\begin{align}
y_t &=\sum_{m=1}^{M}s_{m,t}(\mu_{m,t}+\sigma_{m,t}\varepsilon_{m,t}), \quad \varepsilon_{m,t}\sim\text{IID}(0,1), \label{eq:defeq}\\
\mu_{m,t} &=\varphi_{m,0}+\sum_{i=1}^{p}\varphi_{m,i}y_{t-i}, \quad m=1,...,M, \label{eq:mu_mt}
\end{align}
where $\sigma_{m,t}>0$ are $\mathcal{F}_{t-1}$-measurable, $\varepsilon_{m,t}$ are independent of $\mathcal{F}_{t-1}$, $\varphi_{m,0}\in\mathbb{R}$, $\boldsymbol{\varphi}_m\in\mathbb{S}^p$ (the set $\mathbb{S}^p$ is defined in (\ref{eq:sp})), and $s_{1,t},...,s_{M,t}$ are unobservable regime variables such that for each $t$, exactly one of them takes the value one and the others take the value zero. Given the past of $y_t$, $s_{m,t}$ and $\varepsilon_{m,t}$ are assumed to be conditionally independent, and the conditional probability for regime $m$ occurring at the time $t$ is expressed in terms of the mixing weights $\alpha_{m,t}\equiv \text{Pr}\left(s_{m,t}=1\right|\mathcal{F}_{t-1})$ that satisfy $\sum_{m=1}^{M}\alpha_{m,t}=1$ (for all $t=1,2,...$). Each observation is thus generated by a linear autoregression corresponding to some (unobserved) mixture component $m$ which is selected randomly according to the probabilities determined by the mixing weights.

%In the G-StMAR model, each observation is generated by one of its mixture components. Conditionally on the past, the (unobserved) mixture component is selected according to the probabilities referred to as the mixing weights. The first $M_1$ mixture components are (linear) Gaussian autoregressions and the rest $M_2$ are Student's $t$ autoregressions, yielding a total of $M_1+M_2\equiv M$ mixture components.
The first $M_1$ mixture components are (linear) Gaussian autoregressions and the rest $M_2\equiv M-M_1$ are Student's $t$ autoregressions. Regarding equation (\ref{eq:defeq}), this means that for $m=1,...,M_1$, the terms $\varepsilon_{m,t}$ have standard normal distributions and the variances $\sigma_{m,t}^2$ are constants $\sigma_m^2$. For $m=M_1+1,...,M$, the terms $\varepsilon_{m,t}$ follow the $t$-distribution $t_1\left(0,1,\nu_m+p\right)$ and the variances $\sigma_{m,t}^2$ are as in equation (\ref{eq:sigma_mt}) except that $\boldsymbol{z}_{t-1}$ is replaced with $\boldsymbol{y}_{t-1}=(y_{t-1},...,y_{t-p})$ and the regime specific parameters $\varphi_{m,0},\boldsymbol{\varphi}_m,\sigma_m^2,\nu_m$ are used to define $\mu$ and $\bold{\Gamma}_p$ therein. The component specific conditional means $\mu_{m,t}$ are defined by equation (\ref{eq:mu_mt}) for all the components.

Based on the above specifications, the conditional density function of a G-StMAR model with autoregressive order $p$ is given as 
\begin{equation}\label{gstmar:conddens}
f\left(y_t |\mathcal{F}_{t-1}\right)=\sum_{m=1}^{M_1}\alpha_{m,t}n_1(y_t;\mu_{m,t},\sigma_m^2)+\sum_{m=M_1+1}^{M}\alpha_{m,t} t_1\left(y_t;\mu_{m,t},\sigma_{m,t}^2,\nu_m+p\right),
\end{equation}
where the conditional densities $n_1(y_t;\mu_{m,t},\sigma_m^2)$ and $t_1\left(y_t;\mu_{m,t},\sigma_{m,t}^2,\nu_m+p\right)$ are obtained from the properties of the component processes (using the regime specific parameters). The form of the Student's $t$ density function in (\ref{gstmar:conddens}) is given in online Appendix. The G-StMAR model adds to the class of mixture models introduced by \cite*{Le+Martin+Raftery:1996} and further developed by \cite{Wong+Li:2000, Wong+Li:2001, Wong+Li2:2001}, \cite{Glasbey:2001}, \cite{Lanne+Saikkonen:2003}, and \cite{Wong+Chan+Kam:2009}, to name a few.

In order to specify the mixing weights $\alpha_{m,t}$ in (\ref{gstmar:conddens}), we first define the following function for notational convenience. Let
\begin{equation}
d_m(\boldsymbol{y};\mu_m\mathbf{1}_p,\boldsymbol{\Gamma}_m,\nu_m)=
\left\{\begin{matrix*}[l]
 n_p(\boldsymbol{y};\mu_m\mathbf{1}_p,\boldsymbol{\Gamma}_m), & \text{when} \ m \leq M_1, \\
 t_p(\boldsymbol{y};\mu_m\mathbf{1}_p,\boldsymbol{\Gamma}_m,\nu_m), & \text{when} \ m > M_1,
\end{matrix*}\right.
\end{equation}
where the $p$-dimensional densities $n_p(\boldsymbol{y};\mu_m\mathbf{1}_p,\boldsymbol{\Gamma}_m)$ and $t_p(\boldsymbol{y};\mu_m\mathbf{1}_p,\boldsymbol{\Gamma}_m,\nu_m)$ correspond to the stationary distribution of the $m$th component process (given in the equations (\ref{eq:gausdistr1}) and (\ref{eq:studentdist1})). Denoting $\boldsymbol{y}_{t-1}=(y_{t-1},...,y_{t-p})$, the mixing weights of the G-StMAR model are defined as
\begin{equation}\label{gstmar:alphamt}
\alpha_{m,t}=\frac{\alpha_m d_m(\boldsymbol{y}_{t-1};\mu_m\mathbf{1}_p,\boldsymbol{\Gamma}_m,\nu_m)}{\sum_{n=1}^{M}\alpha_n d_n(\boldsymbol{y}_{t-1};\mu_n\mathbf{1}_p,\boldsymbol{\Gamma}_n,\nu_n)},
\end{equation}
where the parameters $\alpha_1,...,\alpha_M$ satisfy $\sum_{m=1}^M\alpha_m=1$. The mixing weights are thus weighted ratios of densities of the component processes corresponding to the $p$ previous observations. This specific definition of the mixing weights is appealing as it states that an observation is more likely to be generated from a regime with higher relative weighted likelihood. Moreover, it allows the probabilities of each regime occurring to depend on the level, variability, and temporal dependence of the past observations. This is not only convenient for forecasting but it also allows the researcher to associate specific characteristics to different regimes. It turns out that this formulation of the mixing weights also leads to attractive theoretical properties such as fully known stationary distribution of realizations $(y_t,...,y_{t-h})$, $h=0,1,...,p$, and ergodicity of the process. These theoretical properties are formally stated in Theorem \ref{thm:stationarydist} below.

Before stating the theorem, a few notational conventions are provided. We collect the parameters of the G-StMAR model to a $(M(p+3)+M_2-1)\times1$ vector $\boldsymbol{\theta} \equiv (\boldsymbol{\theta}^-,\boldsymbol{\nu})$ where $\boldsymbol{\theta}^-=(\boldsymbol{\vartheta}_1,...,\boldsymbol{\vartheta}_M,\alpha_1,...,\alpha_{M-1})$, $\boldsymbol{\vartheta}_m=(\varphi_{m,0},\boldsymbol{\varphi}_m,\sigma^2_m)$, $\boldsymbol{\varphi}_m=(\varphi_{m,1},...,\varphi_{m,p})$, $m=1,...,M$, and $\boldsymbol{\nu}=(\nu_{M_1+1},...,\nu_M)$. The parameter $\alpha_M$ is omitted because it is obtained from the restriction $\sum_{m=1}^M\alpha_m=1$. The parameter space for the G-StMAR model is
\begin{multline}\label{gstmar:paramspace}
\boldsymbol{\Theta} = \Bigl\lbrace \boldsymbol{\theta}\in \mathbb{R}^{M(2 + p)}\times(0,1)^{M-1}\times (2,\infty)^{M_2} :  \boldsymbol{\varphi}_m\in\mathbb{S}^p, \sigma_m^2>0, \text{ for all } m=1,...,M \Bigr\rbrace
\end{multline}
where the restriction $\nu_m>2$ ($m=M_1+1,...,M$) is made to ensure existence of finite second moments and the set $\mathbb{S}^p$ is as in (\ref{eq:sp}). A G-StMAR model with autoregressive order $p$, $M_1$ GMAR type regimes, and $M_2$ StMAR type regimes is referred to as the G-StMAR($p,M_1,M_2$) model, whenever clarity of the presentation requires. 

\begin{theorem}\label{thm:stationarydist}
Consider the G-StMAR process $y_t$ generated by (\ref{gstmar:conddens}) and (\ref{gstmar:alphamt}) with $\boldsymbol{\theta}\in\boldsymbol{\Theta}$. Then $\boldsymbol{y}_t=(y_t,...,y_{t-p+1})$ ($t=1,2,...$) is a Markov chain on $\mathbb{R}^p$ with a stationary distribution characterized by the density
\begin{equation}
f(\boldsymbol{y};\boldsymbol{\theta})
=\sum_{m=1}^{M_1}\alpha_m n_p(\boldsymbol{y};\mu_m\mathbf{1}_p,\boldsymbol{\Gamma}_m)+\sum_{m=M_1+1}^{M}\alpha_m t_p(\boldsymbol{y};\mu_m\mathbf{1}_p,\boldsymbol{\Gamma}_m,\nu_m).
\end{equation}
Moreover, $\boldsymbol{y}_t$ is ergodic.
\end{theorem}  
The stationary distribution of $\boldsymbol{y}_t$ is a mixture of $p$-dimensional normal and $t$-distributions with constant mixing weights $\alpha_m$. By the well known properties of the normal and the $t$-distribution, all its moments lower than $\min\lbrace\nu_{M_1+1},...,\nu_{M}\rbrace$ exist and are finite. Moreover, as shown in the proof of Theorem \ref{thm:stationarydist}, for any $h=0,1,...,p$, the marginal stationary distribution of the vector $(y_t,..,y_{t-h})$ is also a mixture of normal and $t$-distributions. This gives the parameters $\alpha_m$ an interpretation as the unconditional probabilities for the observation $y_t$ being generated from the $m$th component process. Similarly to the GMAR and the StMAR process, the mean, variance, and first $p$ autocovariances of $y_t$ are thus
\begin{equation}
\text{E}[y_t]\equiv \mu_y =\sum_{m=1}^M\alpha_m\mu_m, \enspace \gamma_j \equiv \sum_{m=1}^M\alpha_m\gamma_{m,j} + \sum_{m=1}^M\alpha_m(\mu_m - \mu_y)^2, \enspace j=0,1,...,p,
\end{equation}
where $\gamma_{m,j}$ is the $j$:th autocovariance of the $m$:th component process. 

The conditional mean and variance of the G-StMAR process are obtained from the definition of the model as 
$
\text{E}[y_t|\mathcal{F}_{t-1}]=\sum_{m=1}^{M}\alpha_{m,t}\mu_{m,t}
$
and
\begin{equation}
\text{Var}(y_t|\mathcal{F}_{t-1})=\sum_{m=1}^{M_1}\alpha_{m,t}\sigma^2_{m}+\sum_{m=M_1+1}^{M}\alpha_{m,t}\sigma^2_{m,t}+\sum_{m=1}^{M}\alpha_{m,t}\left(\mu_{m,t}-\sum_{n=1}^{M}\alpha_{n,t}\mu_{n,t}\right)^2.
\end{equation}
The conditional mean shares a common form with the GMAR model and StMAR model but differs from them in the definition of the mixing weights. The conditional variance includes three components; the first one is related to the conditional variances of the GMAR type components and the second one to the StMAR type components, whereas the third term encapsulates heteroskedasticity caused by variations in the conditional mean.

Notice that the GMAR model \citep{Kalliovirta+Meitz+Saikkonen:2015} can be obtained as a special of the G-StMAR model by setting $M_1=M$ and $M_2=0$, and similarly the StMAR model \citep{Meitz+Preve+Saikkonen:2018} is obtained by setting $M_1=0$ and $M_2=M$. We simply need to drop the corresponding terms from the formulas, and all the definitions and results stated in this and in the next section also hold for to the GMAR and StMAR models individually. However, some theory developed for the GMAR model, such as geometric ergodicity \citep[Theorem A.1]{Kalliovirta+Meitz+Saikkonen:2015}, has not been established for the StMAR and G-StMAR models. The GMAR model also requires less (currently) unverified assumptions than the StMAR and G-StMAR models for concluding asymptotic normality of the maximum likelihood estimator (see \citeauthor{Kalliovirta+Meitz+Saikkonen:2015}, \citeyear{Kalliovirta+Meitz+Saikkonen:2015}, Section 2, \citeauthor{Meitz+Preve+Saikkonen:2018}, \citeyear{Meitz+Preve+Saikkonen:2018}, Theorem 3, and Theorem \ref{thm:mle} of this paper)

\section{Estimation}\label{sec:estim}
Parameters of the G-StMAR model can be estimated with the method of maximum likelihood (ML). Because the stationary distribution of the process is known, the exact log-likelihood function can be used. Suppose the observed time series is $y_{-p+1},...,y_0,y_1,...,y_T$ and that the initial values are stationary. Then the log-likelihood function of the G-StMAR model takes the form
\begin{multline}\label{estimation:loglik}
L(\boldsymbol{\theta})=
\log\left(\sum_{m=1}^{M_1}\alpha_m n_p(\boldsymbol{y}_0;\mu_m\mathbf{1}_p,\boldsymbol{\Gamma}_m)+\sum_{m=M_1+1}^{M}\alpha_m t_p(\boldsymbol{y}_0;\mu_m\mathbf{1}_p,\boldsymbol{\Gamma}_m,\nu_m) \right)+\sum_{t=1}^{T}l_t(\boldsymbol{\theta}),
\end{multline}
where
\begin{equation}\label{estimation:loglik2}
l_t(\boldsymbol{\theta})=\text{log}\left(\sum_{m=1}^{M_1}\alpha_{m,t}n_1(y_t;\mu_{m,t},\sigma_m^2)+\sum_{m=M_1+1}^{M}\alpha_{m,t} t_1\left(y_t;\mu_{m,t},\sigma_{m,t}^2,\nu_m+p\right)\right),
\end{equation}
and the density functions $n_d(\cdot;\cdot)$ and $t_d\left(\cdot;\cdot\right)$ follow the notation described in Section \ref{sec:gstmar}. If stationarity of the initial values seems unreasonable, one can condition on the initial values by dropping the first term on the right hand side of (\ref{estimation:loglik}) and base the estimation on the resulting conditional log-likelihood function.

In what follows, we assume estimation based on the conditional log-likelihood function $L_T^{(c)}(\boldsymbol{\theta})=T^{-1}\sum_{t=1}^Tl_t(\boldsymbol{\theta})$, i.e., that the ML estimator $\hat{\boldsymbol{\theta}}_T$ maximizes $L_T^{(c)}(\boldsymbol{\theta})$. We have scaled the conditional log-likelihood function with the sample size $T$ so that the notation is consistent with the referred literature. 

To investigate the asymptotic properties of the ML estimator $\hat{\boldsymbol{\theta}}_T$, the parameter space $\boldsymbol{\Theta}$ given in (\ref{gstmar:paramspace}) needs to be restricted in a way that guarantees identification of the parameters. This amounts to requiring that components of the G-StMAR model cannot be "relabelled" so that one ends up with the same model with different parameter vector; that is,
\begin{multline}\label{gstmar:identcond}
\alpha_1>\cdots>\alpha_ {M_1}>0, \enspace \alpha_{M_1+1}>\cdots>\alpha_ {M}>0, \text{ and }\boldsymbol{\vartheta}_i=\boldsymbol{\vartheta}_j \text{ only if some of the conditions}\\ (1) \enspace 1\leq i=j\leq M,
\enspace (2) \enspace i\leq M_1 < j, \enspace (3) \enspace i,j>M_1 \text{ and } \nu_i\neq \nu_j \text{ is satisfied.}\quad\quad\quad\;
\end{multline}
The restrictions required to establish asymptotic properties of the ML estimator are summarized in the following assumption.
\begin{assumption}\label{assump:param}
The true parameter value $\boldsymbol{\theta}_0$ is an interior point of $\boldsymbol{\bar{\Theta}}$ which is a compact subset of $\lbrace \boldsymbol{\theta}\in\boldsymbol{\Theta}:(\ref{gstmar:identcond}) \text{ holds} \rbrace$.
\end{assumption}

Asymptotic properties of the ML estimator under the conventional high-level conditions are stated in the following theorem (which is similar to Theorem 3 in \cite{Meitz+Preve+Saikkonen:2018} on the ML estimator of the StMAR model). Denote $\mathcal{I}(\boldsymbol{\theta}) = \text{E}\big[\frac{\partial l_t(\boldsymbol{\theta})}{\partial\boldsymbol{\theta}}\frac{\partial l_t(\boldsymbol{\theta})}{\partial\boldsymbol{\theta}'} \big]$ and $\mathcal{J}(\boldsymbol{\theta})=\text{E}\big[\frac{\partial^2 l_t(\boldsymbol{\theta})}{\partial\boldsymbol{\theta}\partial\boldsymbol{\theta}'}\big]$.
\begin{theorem}\label{thm:mle}
Suppose that $y_t$ are generated by the stationary and ergodic G-StMAR process of Theorem \ref{thm:stationarydist} and that Assumption \ref{assump:param} holds. Then $\hat{\boldsymbol{\theta}}_T$ is strongly consistent, i.e., $\hat{\boldsymbol{\theta}}_T \rightarrow \boldsymbol{\theta}_0$ almost surely. Suppose further that (i) $T^{1/2}\frac{\partial}{\partial\boldsymbol{\theta}}L_T^{(c)}(\boldsymbol{\theta}_0)\overset{d}{\rightarrow}N(0,\mathcal{I}(\boldsymbol{\theta}_0))$ with $\mathcal{I}(\boldsymbol{\theta}_0)$ finite and positive definite, (ii) $\mathcal{J}(\boldsymbol{\theta}_0)=-\mathcal{I}(\boldsymbol{\theta}_0)$, and (iii) $\text{E}\big[\sup_{\boldsymbol{\theta}\in \boldsymbol{\bar{\Theta}}_0}\big| \frac{\partial^2l_t(\boldsymbol{\theta})}{\partial\boldsymbol{\theta}\partial\boldsymbol{\theta}'} \big| \big] <\infty$ for some $\boldsymbol{\bar{\Theta}}_0$, compact convex set contained in the interior of $\boldsymbol{\bar{\Theta}}$ that has $\boldsymbol{\theta}_0$ as an interior point. Then $T^{1/2}(\hat{\boldsymbol{\theta}}_T - \boldsymbol{\theta}_0) \overset{d}{\rightarrow} N(0,-\mathcal{J}(\boldsymbol{\theta}_0)^{-1})$.
\end{theorem}
%The conditions (i)-(iii) of Theorem \ref{thm:mle} are standard for establishing the limiting distribution of an ML estimator. The first condition states that when evaluated at $\boldsymbol{\theta}_0$, the score vector fulfils a central limit theorem, and that the information matrix is positive definite. The second condition is the information matrix equality, and the third one is required to conclude that the Hessian matrix converges uniformly in a neighbourhood of the true parameter value.
If one is willing to assume validity of the conditions (i)-(iii) of Theorem \ref{thm:mle}, the ML estimator $\hat{\boldsymbol{\theta}}_T$ has the conventional limiting distribution, implying that approximative standard errors for the estimates are obtained as usual. Moreover, standard likelihood based tests are applicable as long as the orders $M_1$ and $M_2$ are correctly specified. If $M_1$ or $M_2$ is chosen too large, some of the parameters are not identified causing the result of Theorem \ref{thm:mle} to break down. This particularly happens when one tests for the number of regimes as the null hypothesis would imply that some regime is reduced from the model\footnote{\cite{Meitz+Saikkonen:2017} have, however, recently developed such tests for mixture models with Gaussian conditional densities.} \citep[see the related discussion in][Section 3.3.2]{Kalliovirta+Meitz+Saikkonen:2015}. 
Similar caution also applies for testing whether a regime is of the GMAR type against the alternative that it is of the StMAR type, as under the null hypothesis $\nu_m=\infty$ for the StMAR type regime $m$ being tested, violating Assumption \ref{assump:param}. Numerical consequences of the weak identification of very large degrees of freedom parameters are briefly discussed in Section \ref{sec:building}. 

\subsection{Two-phase maximum likelihood estimation}\label{sec:twophase}
Finding the ML estimates amounts to maximizing the log-likelihood function (\ref{estimation:loglik}) over the high dimensional parameter space (\ref{gstmar:paramspace}) satisfying several constraints. Due to the complexity of the log-likelihood function, finding an analytical solution is infeasible, so numerical optimization methods are required. The EM algorithm \citep{Redner+Walker:1984} has been a popular choice for estimating mixture models \cite[e.g.][and \citealp{Wong+Chan+Kam:2009}]{Wong+Li:2000, Wong+Li:2001, Wong+Li2:2001} as it is suitable for problems where all the data relevant to estimation is not observed (for mixture models that is the origin of each observation; in our case, the random variables $s_{1,t},...,s_{M,t}$ in (\ref{eq:defeq})). For the G-StMAR model the EM algorithm is not, however, particularly useful because in each maximization step one faces a new optimization problem that is not much simpler than the original one. This is because in the G-StMAR model the mixing weights also depend on the AR parameters (in a complex way). Conventional gradient based algorithms, on the other hand, tend to converge to some local maximum near the starting point, making them generally insufficient for maximizing multimodal objective functions such as (\ref{estimation:loglik}) that require thorough exploration of the parameter space.

Several optimization algorithms capable of escaping from local maxima have been proposed for maximization of complicated multimodal objective functions. Such robust methods, which include simulated annealing and the genetic algorithm (see, e.g., \citeauthor*{Goffe+Ferrier+Rogers:1994}, \citeyear{Goffe+Ferrier+Rogers:1994} and \citeauthor{Dorsey+Mayer:1995}, \citeyear{Dorsey+Mayer:1995}), often perform well but they are computationally heavy and tend to converge slowly when near the global maximum point \citep[see the discussion in][Section 3]{Dorsey+Mayer:1995}. Following \cite{Dorsey+Mayer:1995} \citep[and][]{Meitz+Preve+Saikkonen:2018,Meitz+Preve+Saikkonen2:2018}, we hence suggest employing a hybrid estimation procedure where a genetic algorithm is used to find starting values for a gradient based method which then accurately converges to a nearby local maximum or saddle point. %(\citeauthor{Meitz+Preve+Saikkonen:2018} \citeyear{Meitz+Preve+Saikkonen:2018} also utilized a similar procedure).

Even with the two-phase estimation procedure, parameters of the G-StMAR model can be challenging to estimate. We have therefore accompanied this paper with the CRAN distributed R package "uGMAR" \citep{uGMAR} in which the genetic algorithm has been modified to improve its performance.\footnote{In addition to the G-StMAR model, uGMAR also accomodates the GMAR and StMAR models.} Brief descriptions of the employed genetic algorithm and its modifications are given in an Appendix. After running the genetic algorithm, the estimation is finalized with a variable metric algorithm \cite[algorithm 21, implemented by \citealp{Rnormal}]{Nash:1990} using central difference approximation for the gradient of the log-likelihood function. Because of the presence of multiple local maxima, a (sometimes large) number of estimation rounds should be performed to obtain reliable results, for which uGMAR makes use of parallel computing to shorten the estimation time.

\section{Building a G-StMAR Model}\label{sec:building}
In empirical applications, building a G-StMAR model amounts to finding a suitable autoregressive order $p$, the number of GMAR type regimes $M_1$, and the number of StMAR type regimes $M_2$. Different strategies for choosing the number of each type of regimes may be considered depending on the application. We propose a simple model selection procedure which takes advantage of the observation that the G-StMAR model is a limiting case of the StMAR model\footnote{The definition of the StMAR($p,M$) model is technically the same as of the G-StMAR($p,0,M$) model.}.

It is easy to see that the linear Gaussian autoregression defined in Section \ref{sec:gstmar_gausstud} is obtained as a limiting case of the Student's $t$ autoregression with the degrees of freedom parameter tending to infinity. As the mixing weights (\ref{gstmar:alphamt}) are weighted ratios of the component process densities, it then follows that the G-StMAR($p,M_1,M_2$) model is obtained as a limiting case of a StMAR($p,M$) model with the parameters $\nu_1,...,\nu_{M_1}$ limiting to infinity. Consequently, if a StMAR($p,M$) model is fitted to data generated by a G-StMAR($p,M_1,M-M_1$) process, then asymptotically, the $M_1$ regimes of the fitted StMAR model are expected to get large degrees of freedom estimates. We therefore suggest building a G-StMAR model by first finding a suitable StMAR model, and then estimating the appropriate G-StMAR model if the fitted StMAR model contains large degrees of freedom estimates. A StMAR model can be specified, for example, by using information criteria together with quantile residual diagnostics \citep[see, e.g.,][]{Kalliovirta:2012}. %Notice that this procedure for building a G-StMAR model does not necessarily lead to the model that minimizes an information criterion, say, Akaike's information criterion among StMAR and G-StMAR models with some maximum orders for $p$ and $M$.

Overly large degrees of freedom estimates in a StMAR model are redundant but their weak identification also causes several inconveniences in numerical analysis of the model. They lead to nearly numerically singular Hessian matrix of the log-likelihood function when evaluated at the estimate, making the approximate standard errors often unavailable. Weakly identified degrees of freedom parameters also cause inconvenience in quantile residual based model diagnostics. In particular, the quantile residual tests proposed by \cite{Kalliovirta:2012} require a positive definite approximation of the Hessian matrix (evaluated the ML estimate). The tests are thus not applicable for StMAR models with too large degrees of freedom estimates, whereas they are for the corresponding G-StMAR models. Applicability of \citeauthor{Kalliovirta:2012}'s (\citeyear{Kalliovirta:2012}) tests, which take into account the uncertainty caused by estimation of the parameters, might have consequences in model selection when sheer graphical analysis of the quantile residuals fails to reveal inadequacies. We demonstrate such a case in the empirical application.

\section{Empirical application}\label{sec:empexample} 
\begin{figure}[t]
    \centerline{\includegraphics[scale=0.652]{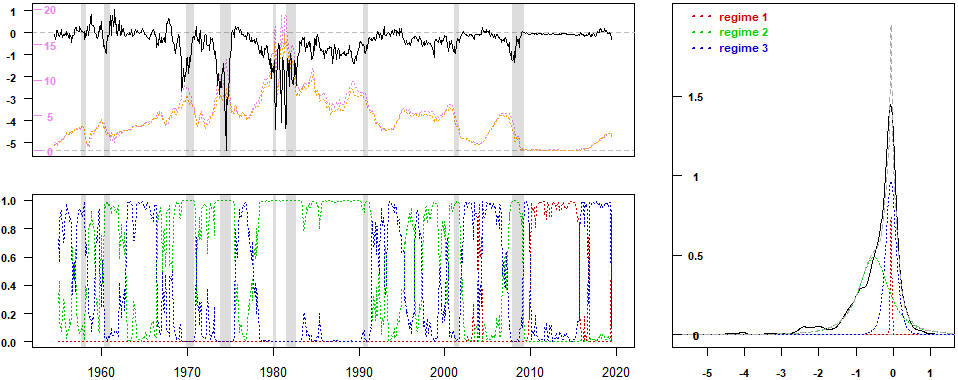}}
    \caption{On top left, monthly U.S. 3-month Treasury bill secondary market rate minus effective federal funds rate (black solid line), the 3-month Treasury bill secondary market rate (orange dotted line), and the effective federal funds rate (violet dotted line). On bottom left, the mixing weights implied by the G-StMAR($5,1,2$) model fitted to the interest rate spread series. The shaded areas indicate the periods of (NBER based) U.S. recessions. On right, a Gaussian kernel density estimate of the interest rate spread (black solid line), the mixture density implied by the fitted G-StMAR($5,1,2$) model (grey dashed line), and the regime densities (blue, green, and red dotted lines).}
\label{fig:mod53}
\end{figure}

We consider the monthly U.S. interest rate spread between the 3-month Treasury bill (TB) secondary market rate and the effective federal funds (FF) rate, covering the period from 1954VII to 2019VII (781 observations). The series is plotted in Figure \ref{fig:mod53} (top left) along with the 3-month TB and FF rates, and with the shaded areas indicating the periods of (NBER based) U.S. recessions. All the data were taken from the Federal Reserve Bank of St. Louis database.

Treasury bills are short-term pure discount bonds which are backed by the U.S. government and therefore generally considered to be almost free from default-risk. The effective federal funds rate is the averaged rate at which depository institutes loan federal funds to each other overnight. The overnight FF lending agreements are one of the most liquid financial asset, but unlike TBs, they are subject to a notable default-risk. 
The relationship between TB and FF rates has been studied, among others, by \cite{Simon:1990} and \cite{Sarno+Thornton:2003}, while \cite{Kishor+Marfatia:2013} examine the relationship between TB and FF futures rate. 

According to term structure theory, a long-term interest rate should reflect the current and expected future short-term rates, and also perceptions of risk and liquidity in the form of (possibly time varying) premium. \cite{Simon:1990} studied the predictive power of the weekly spread between the 3-month TB and FF rates on the future levels of the FF rate in 1972-1987. He argued that the current and expected future FF rates affect the spread between the TB and FF rates through the repurchase agreement (repo) market\footnote{In a repo, the \textit{borrower} sells a security to the \textit{lender} and agrees to repurchase it in the future (often in the next day). Effectively, repos function similarly to collateralized loans. See \cite{Baklanova+Copeland+McCaughrin:2015} for an overview of the U.S. repo market.} because repos are closely linked to the FF rate, and corporations with funds to invest can buy TBs alternatively to investing in consecutive overnight repos. TB rates are linked to the FF rates also because security dealers finance the bulk of their TB inventories in the repo market, which is closely tied to the FF market. Furthermore, when trust in solidity of the banking system weakens, the increased demand for safety lowers TB rates relatively to FF rates. \cite{Simon:1990} accounted for this by employing the spread between the 3-month Eurodollar time deposit\footnote{Eurodollar time deposit is a U.S. dollar-denominated deposit at a bank outside the U.S. with a fixed maturity.} and TB rates as a risk premium for bank safety. He found that the spread between the 3-month TB and FF rate had significant predictive power on future levels of the FF rate in the volatile nonborrowed reserves operating period (late 1979 - late 1982) but less or none in the other subperiods.

\cite{Sarno+Thornton:2003} identified an error correction model (ECM) between the daily 3-month TB and effective FF rate (covering the period from 1974 to 1999) and showed that their ECM, which allows for asymmetries and nonlinearities, outperforms the alternative of a linear ECM. One of their main findings was that the FF rate (which is controlled by the Fed) seems to adjust to the TB rate and not vice versa, supporting the hypothesis that the market anticipates changes in the FF rate, moving the TB rate in advance. Moreover, it appears that the adjustment speed depends on the sign and size of the deviation from the long-run equilibrium. \cite{Sarno+Thornton:2003} argued that although there has been a number of procedural changes affecting predictability of the FF rate, their results implicate that the changes have been statistically unimportant. Furthermore, their robustness checks indicate that their findings on the adjustments from disequilibria also hold for monthly data. Variations and asymmetries in the adjustment speed, on the other hand, indicate that the dynamics of the spread between the TB and FF rates might fluctuate along with the level of the spread. This suggests that a mixture model, such as the G-StMAR model, which is able encapsulate such behaviour could be an appropriate choice of model.

\cite{Kishor+Marfatia:2013} argued that the results in \cite{Sarno+Thornton:2003} are not very surprising since the effective FF rate always tends to revert back to the FF target rate, and it does not incorporate markets expectations of the changes in the future FF rate. To get around that, they studied the relationship between the 3-month TB rate and the 1-month FF futures rate which does incorporate information about market's anticipations on the future FF rate. They fitted a linear ECM to a daily series from 1989 to 2008, and found that the TB rate and the FF futures rate both seem to move to correct a short-run disequilibrium.

Interestingly, the spread between the 3-month TB rate and the effective FF rate is most of the time (covered in our sample period) negative. \cite{Sarno+Thornton:2003} made a similar observation for their daily series and suggested that only a small fraction of the negative difference could be attributed to the low default-risk of TBs, but that a more plausible explanation is that the interest on TBs is exempt from some local and state taxes. As the smaller taxes have larger effect on paid net interest (relative to interest paid on federal funds) when the interest rates are higher, some movements of the spread could be partially caused by the differences in taxation.

\subsection{Estimation and model selection}
We employ the method of maximum likelihood based on the exact log-likelihood function for estimating the parameters of the considered models. Adequacy of the estimated models is examined using quantile residual diagnostics in the framework presented in \cite{Kalliovirta:2012}. The quantile residuals of a correctly specified G-StMAR model are asymptotically independent with standard normal distributions \citep[Lemma 2.1]{Kalliovirta:2012}, so they can be used for graphical analysis in a similar fashion to conventional Pearson's residuals. In addition to graphical analysis of the quantile residuals, we perform \citeauthor{Kalliovirta:2012}'s (\citeyear{Kalliovirta:2012}) asymptotic tests (which take into account the uncertainty caused by estimation of the parameters) for testing normality, autocorrelation, and conditional heteroskedasticity of the quantile residuals. The estimation, quantile residual diagnostics, and other numerical analysis of the models is conducted using the R package uGMAR \citep{uGMAR} which is available through the CRAN repository.\footnote{There is also Matlab code available for the StMAR model in the form of StMAR MATLAB Toolbox by \cite{Meitz+Preve+Saikkonen2:2018}.} uGMAR estimates the model parameters using the two-phase procedure described in Section \ref{sec:twophase}.
%by first using a genetic algorithm \citep[see, e.g.,][]{Dorsey+Mayer:1995} to find starting values for a variable metric algorithm \citep[algorithm 21]{Nash:1990} which then convergences to the nearby local maxima.

%We employed information criteria for preliminary model selection and found that the
Following the model selection procedure described in Section \ref{sec:building}, we started by finding a suitable StMAR model. First, we estimated the StMAR($p, M$) model with one mixture component, $M = 1$, and autoregressive orders $p=1,...,24$ and found that the order $p = 6$ yields the largest likelihood. Adequacy of the StMAR($6, 1$) model was clearly rejected by the quantile residual tests (see Table 2), so we estimated the StMAR($p, M$) models with orders $p = 1,...,6$ and $M = 2,3$. The order $(p,M)=(5,2)$ minimized the Schwarz-Bayesian (BIC) and the Hannan-Quinn (HQIC) information criteria, whereas the Akaike's information criterion (AIC) was minimized by the order $(p,M)=(5,3)$. Inappropriate estimates extremely near the border of the stationarity region were discarded as they are not solutions of interest (but maximize the likelihood for rather a technical reason), so in such cases the next-best local maximum of the log-likelihood function was considered instead. In both the StMAR($5,2$) and the StMAR($5,3$) model, a very large degrees of freedom estimate for one regime was obtained (approximately $99000$ and $95000$, respectively), so we estimated the corresponding G-StMAR($5,1,1$) and G-StMAR($5,1,2$) models. Removing the weakly identified degrees of freedom parameters by switching to the G-StMAR models enabled us to compute approximate standard errors of the estimates and to calculate \citeauthor{Kalliovirta:2012}'s (\citeyear{Kalliovirta:2012}) test statistics (see Section \ref{sec:building}). The values of the information criteria are reported in Table \ref{tab:qrtests} and the parameter estimates of the G-StMAR models are reported in Table \ref{tab:MLestimates} with the approximate standard errors for the estimates in brackets. 

Estimates regarding the GMAR type regime are quite similar for the two G-StMAR models, and their standard errors are relatively large. This is because for both of the models the GMAR type regime mainly occurs in the period of near-zero interest rates after 2008 and there are hence only few observations from that regime (regime 1 in Figure \ref{fig:mod53}, bottom left, which displays the mixing weights of the G-StMAR($5,1,2$) model; the mixing weights of the G-StMAR($5,1,1$) model are not shown). The three zeros in the variance parameter estimates (and in their standard errors) signify that the estimates (and their standard errors) round to zero in three digits accuracy\footnote{More accurate values for the ML estimate of $\sigma^2_1$ and its standard error are $3.237\times 10^{-4}$ and $6.884\times 10^{-5}$ for the G-StMAR($5,1,1$) model, $3.070\times 10^{-4}$ and $6.092\times 10^{-5}$ for the G-StMAR($5,1,2$) model, and $3.593\times 10^{-4}$ and $5.552\times 10^{-5}$ for the G-StMAR($5,1,2)^r$ model, respectively.}, implying that the GMAR type regime exhibits very low variability (conditionally and unconditionally). The small mixing weight parameter estimates, interpreted as the unconditional probability for the GMAR type regime occurring, reflect the observation that eras of such a low variability have been rare in the sample period. Also, a remarkably large standard error for the second regime's variance parameter sticks out for both of the models. Examination of the profile log-likelihood functions (not shown) does not, however, reveal anything notable.

Since the AR parameter estimates for the G-StMAR($5,1,2$) model are somewhat similar in all regimes, we estimated a StMAR($5,3$) model with the AR parameters restricted to be the same in all regimes, allowing for changes in the level, variability, and kurtosis only. The degrees of freedom estimate for one regime was very large (approximately $97000$), so we estimated the corresponding restricted G-StMAR model which we refer to as the G-StMAR($5,1,2)^r$ model. The parameter estimates of this model are also presented in Table \ref{tab:MLestimates} with the related statistics, and the values of the information criteria in Table \ref{tab:qrtests}. The standard errors of the AR parameters are notably smaller than in the non-restricted models because the AR parameters are common for all the regimes.

\begin{table}
\centering
\small
\renewcommand{\arraystretch}{0.95}
\begin{tabular}{c S[table-format=-1.3] @{\hspace{0.1cm}} S S[table-format=-1.3] @{\hspace{0.1cm}} S S[table-format=-1.3] @{\hspace{0.1cm}} S}
 & \multicolumn{2}{c}{G-StMAR($5,1,1$)} & \multicolumn{2}{c}{G-StMAR($5,1,2$)} & \multicolumn{2}{c}{G-StMAR($5,1,2)^r$}\\
\hline \\[-1.5ex]
 $\varphi_{1,0}$ & -0.011 & (0.010) & -0.013 & (0.009) & -0.007 & (0.002) \\ 
 $\varphi_{1,1}$ & 0.587  & (0.129) & 0.580  & (0.124) & 0.782  & (0.037) \\  
 $\varphi_{1,2}$ & -0.049 & (0.168) & -0.079 & (0.163) & -0.058 & (0.050) \\
 $\varphi_{1,3}$ & 0.041  & (0.140) & 0.042  & (0.136) & 0.134  & (0.050) \\
 $\varphi_{1,4}$ & 0.006  & (0.142) & 0.006  & (0.141) & -0.040 & (0.052) \\
 $\varphi_{1,5}$ & 0.224  & (0.128) & 0.209  & (0.132) & 0.036  & (0.042) \\
 $\sigma_{1}^2$  & 0.000  & (0.000) & 0.000  & (0.000) & 0.000  & (0.000) \\
 $\alpha_{1}$    & 0.029  & (0.021) & 0.043  & (0.035) & 0.035  & (0.025) \\[1.2ex]
 $\mu_{1}$       & -0.056 &         & -0.055 &         & -0.048 & \\
 $\gamma_{1,0}$  & 0.001  &         & 0.001  &         & 0.001  & \\[1.5ex]

 $\varphi_{2,0}$ & -0.009 & (0.005) & -0.066 & (0.025) & -0.079 & (0.025) \\
 $\varphi_{2,1}$ & 0.821  & (0.040) & 0.845  & (0.055) &        & \\
 $\varphi_{2,2}$ & -0.051 & (0.053) & -0.038 & (0.076) &        & \\
 $\varphi_{2,3}$ & 0.153  & (0.053) & 0.127  & (0.075) &        & \\
 $\varphi_{2,4}$ & -0.052 & (0.055) & -0.134 & (0.077) &        & \\
 $\varphi_{2,5}$ & 0.045  & (0.042) & 0.073  & (0.058) &        & \\
 $\sigma_{2}^2$  & 4.806  & (18.779) & 0.541 & (2.052) & 0.256  & (0.374) \\
 $\nu_{2}$       & 2.007  & (0.026) & 2.196  & (0.801) & 2.499  & (0.872) \\
 $\alpha_{2}$    &        &         & 0.592  & (0.132) & 0.600  & (0.141) \\[1.2ex]
 $\mu_{2}$       & -0.110 &         & -0.519 &         & -0.541 & \\
 $\gamma_{2,0}$  & 24.449 &         & 2.109  &         & 0.802  & \\[1.5ex]
 
 $\varphi_{3,0}$ &        &         & -0.011 & (0.005) & -0.011 & (0.005) \\
 $\varphi_{3,1}$ &        &         & 0.720  & (0.069) &        & \\
 $\varphi_{3,2}$ &        &         & -0.082 & (0.090) &        & \\
 $\varphi_{3,3}$ &        &         & 0.151  & (0.090) &        & \\
 $\varphi_{3,4}$ &        &         & 0.087  & (0.098) &        & \\
 $\varphi_{3,5}$ &        &         & -0.062 & (0.085) &        & \\
 $\sigma_{3}^2$  &        &         & 0.015  & (0.011) & 0.015  & (0.013) \\
 $\nu_{3}$       &        &         & 4.320  & (2.951) & 4.778  & (4.511) \\[1.2ex]
 $\mu_{3}$       &        &         & -0.059 &         & -0.074 & \\
 $\gamma_{3,0}$  &        &         & 0.038  &         & 0.048  & \\[1.5ex]
 
$\mu_y$           & -0.108 &         & -0.331 &         & -0.353 & \\ 
$\gamma_0$      & 23.744 &         & 1.313  &         & 0.552  & \\[2ex]
 
 $L(\hat{\boldsymbol{\theta}})$ & 309.165 & & 322.121 & & 314.016 & \\    
\end{tabular}
\caption{Maximum likelihood estimates of the G-StMAR($5,1,1$), the G-StMAR($5,1,2$), and the restricted G-StMAR($5,1,2)^r$ model based on the exact log-likelihood function, with approximate standard errors for the estimates presented in the brackets. The statistics $\mu_{m}$ and $\gamma_{m,0}$, $m=1,2,3$, are the stationary mean and variance of each regime, respectively. Likewise, the statistics $\mu_y$ and $\gamma_0$ are the stationary mean and variance of the process. The maximized log-likelihoods for each model are presented in the bottom row of the table.} %More accurate estimates for $\sigma^2_1$ and its standard error are $3.237\times 10^{-4}$ and $6.884\times 10^{-5}$ for the G-StMAR($5,1,1$) model, $3.070\times 10^{-4}$ and $6.092\times 10^{-5}$ for the G-StMAR($5,1,2$) model, and $3.593\times 10^{-4}$ and $5.552\times 10^{-5}$ for the G-StMAR($5,1,2)^r$ model, respectively.}
\label{tab:MLestimates}
\end{table}

\begin{figure}[p]
    \centerline{\includegraphics[scale=0.665]{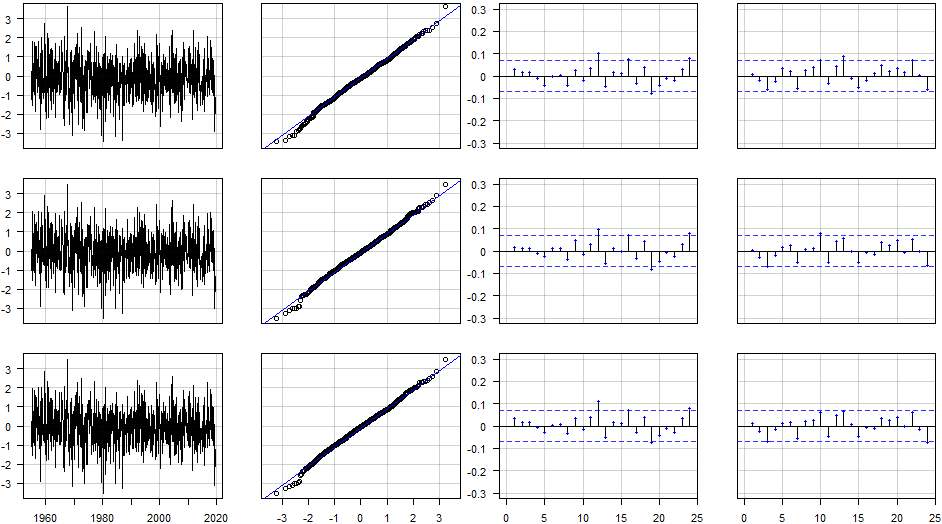}}
    \caption{Graphical quantile residual diagnostics for the models presented in Table \ref{tab:MLestimates}. The top row is for the G-StMAR($5,1,1$) model, the middle row is for the G-StMAR($5,1,2$) model, and the bottom row is for the G-StMAR($5,1,2)^r$ model. The first column presents the time series, the second column the normal quantile plot, and third column the autocorrelation function of the quantile residuals. The fourth column presents the autocorrelation function of the squared quantile residuals. The blue solid line in the quantile plots displays the theoretical quantiles, and the blue dashed lines in the autocorrelation function plots are the 95\% bounds $\pm 1.96/\sqrt{T}$ ($T=776$ as the first $p$ values are the initial values) for autocorrelations of an IID sequence which are presented to give an approximate perception of the magnitude of the sample autocorrelations.}
\label{fig:qrplot}
\end{figure}

\begin{table}[p]
\centering
\renewcommand{\arraystretch}{1}
\hspace*{-0.4cm}
\small\addtolength{\tabcolsep}{-4pt}
\begin{tabular}{c c @{\hspace{0.1cm}} c c c c @{\hspace{0.5cm}} c c c c @{\hspace{0.4cm}} c c c}
  & Normality\; & \multicolumn{4}{c}{Autocorrelation\quad\quad\;} & \multicolumn{4}{c}{Cond. h.skedasticity\quad\quad} & AIC & HQIC & BIC \\
 \hline\\[-2.5ex]
 Number of lags &  & $1$ & $3$ & $6$ & $12$ & $1$ & $3$ & $6$ & $12$ &   &   &   \\
 \hline\\[-2.5ex]
StMAR(6,1)   & $\mathbf{0.00}$ & $\mathbf{0.00}$ & $\mathbf{0.00}$ & $\mathbf{0.00}$ & $\mathbf{0.00}$ & $\mathbf{0.00}$ & $\mathbf{0.00}$ & $\mathbf{0.00}$ & $\mathbf{0.00}$ & $-538$ & $-521$ & $-495$ \\   
 G-StMAR($5,1,1$) & $\mathbf{0.00}$ & $\mathbf{0.01}$ & $\mathbf{0.01}$ & $0.01$ & $\mathbf{0.00}$ & $0.46$ & $0.38$ & $0.23$ & $\mathbf{0.00}$ & $-586$ & $-558$ & $-512$ \\ 
 G-StMAR($5,1,2$) & $\mathbf{0.00}$ & $0.18$ & $0.40$ & $0.57$ & $0.16$ & $0.82$ & $0.07$ & $0.18$ & $0.02$ & $-594$ & $-549$ &  $-478$ \\
 $\enspace$G-StMAR($5,1,2)^r$ & $\mathbf{0.00}$ & $0.02$ & $0.07$ & $0.13$ & $0.03$ & $0.68$ & $0.07$ & $0.24$ & $0.03$ & $-598$ & $-571$ & $-528$
\end{tabular}
\caption{The $p$-values obtained from the \citeauthor{Kalliovirta:2012}'s (\citeyear{Kalliovirta:2012}) quantile residual tests, testing for normality, autocorrelation, and conditional heteroskedasticity of the quantile residuals. The $p$-values smaller than $0.01$ are bolded. In order to improve size properties of the tests, we employed the simulation procedure proposed by \cite{Kalliovirta:2012} using samples of length $500000$.}
\label{tab:qrtests}
\end{table}

Figure \ref{fig:qrplot} presents the time series, normal quantile plot, and the sample autocorrelation function of the quantile residuals, and the sample autocorrelation function of the squared quantile residuals for the G-StMAR models presented in Table \ref{tab:MLestimates}. Graphical analysis of the quantile residuals does not show significant signs of inadequacy for any of the models. A slightly too fat lower tail in the quantile residuals' distributions and somewhat large, approximately $0.1$, sample autocorrelation at lag $12$ sticks out for each of the three models, however. 

In order to further study adequacy of the models, we employed \citeauthor{Kalliovirta:2012}'s (\citeyear{Kalliovirta:2012}) tests, and tested for normality, autocorrelation, and conditional heteroskedasticity of the quantile residuals, taking into account $1,3,6$, and $12$ lags in the autocorrelation and heteroskedasticity tests. The $p$-values obtained from the tests are reported in Table \ref{tab:qrtests}. The normality test rejects for all the three models at $1\%$ level of significance, possibly because of the fat lower tails in the quantile residuals' distributions. More interestingly, despite the similarities in the graphical analysis, the autocorrelation tests unambiguously reject adequacy of the G-StMAR($5,1,1$) model, whereas the $p$-values are reasonable for the G-StMAR($5,1,2$) model which also passes the heteroskedasticity tests. The $p$-values for the autocorrelation tests are rather small also for the restricted G-StMAR($5,1,2)^r$ model, which is preferred by the information criteria, showing some evidence of inadequacy. We therefore prefer the unrestricted G-StMAR($5,1,2$) model whose overall adequacy seems quite satisfactory. The fact that the restricted model has information criteria values superior to the unrestricted models, however, suggests that imposing the autocorrelation structure to be the same for all
regimes would also be a reasonable modelling choice.\footnote{For comparison, we also estimated the GMAR($p,M$) model with orders $p=1,...,6$ and $M=1,...,4$. The values of the information criteria were, however, found inferior to our G-StMAR models, with the GMAR($3,4$) model minimizing BIC ($-432$) and the GMAR($5,4$) model minimizing HQIC ($-517$) and AIC ($-572$).}
%The superior information criteria values, however, suggest that imposing the autocorrelation structure to be the same for all regimes would also be a reasonable modelling choice.

\subsection{Discussion}
Our model selection procedure led to the (unrestricted) G-StMAR($5,1,2$) model which  identifies three statistical regimes for the spread between the 3-month TB secondary market rate and the effective FF rate. The mixing weights of the model are presented in Figure \ref{fig:mod53} (bottom left) along with the interest rate spread series (top left). The GMAR type regime (red)  dominates the period of near-zero interest rates occurring after 2008, where also the spread stays close to zero and exhibits very low variability. The second regime (green) identifies periods of high variability and low mean, spanning through most of the recessions, whereas the third regime (blue) often occurs\footnote{By a regime occurring at a point of time we mean that according to the estimated mixing weights, the process generated an observation from that regime with a probability close to one.} after the recessions when the spread moderately varies around zero. These characteristics of the regimes are also highlighted in Figure \ref{fig:mod53} (right) where a kernel density estimate of the spread (black solid line) is presented with the model implied density (grey dashed line) and the regime densities (red, green, and blue dotted lines; regime densities are multiplied by the mixing weight parameter estimates $\alpha_m$, $m=1,2,3$). The model implied density matches fairly well to the skewed distribution of the observations, but peakiness of the distribution seems a bit exaggerated and the lower tail is not fat enough.

Based on our G-StMAR($5,1,2$) model, the regime specific unconditional mean of the spread varies from the $-0.06$ \%-units of the first (GMAR type) and third regime to the $-0.52$ \%-units of the second regime, with each regime regularly occurring for several consecutive months. 
As the second regime dominates during most of the recessions, and also often occurs before the recessions when the interest rates are relatively high, it seems plausible that part of the larger negative mean is explained by expectations of a decrease in the near-future FF rate. The third regime, on the other hand, mostly occurs after the recessions when the interest rates seem relatively low, possibly indicating that the larger mean of the regime could be related to the lack of expected decreases in the FF rate. These findings are consistent with \cite{Sarno+Thornton:2003} who found that the FF rate corrects disequilibriums from the long-run relationship, supporting the hypothesis that market's anticipations in the future movements of the FF rate are reflected in the TB rate.

\cite{Sarno+Thornton:2003} also found that the adjustment speed of FF rate towards the long-run equilibrium depends on the sign and size of the deviation. Specifically, FF rate below the long-run trend or larger deviation implies faster adjustment, suggesting that too high values of the spread would be corrected faster than too low values. This might partially explain why the low mean second regime usually occurs when the interest rates are declining, but a rise in the FF rate is not always accompanied with a switch to the higher mean third regime. Another possibility is that market's predictions on the future movements of the FF rate are sometimes rather poor or a premium has an increased effect on the opposite direction. During the savings and loan crisis in 80's and 90's, increased preference for the safety of TBs would seem like a plausible partial explanation for the moderately negative spread despite of the mainly increasing FF rate from late 1986 to early 1989.

Overall, the three statistical regimes of our G-StMAR model identify three economic regimes, with the first regime dominating the period in which the movements of the interest rates are limited by the zero lower bound. The second regime arguably occurs often when the market anticipates decreases in the FF rate or possibly has increased preferences for the safety of the almost default-risk free TBs. The third regime seems to mostly occur at times when the Fed is arguably not expected to significantly decrease the FF rate target (because the recession has already passed and the interest rates are relatively low). %Although our G-StMAR model explains the autocorrelation structure and variations of the spread fairly adequately, this empirical application is not intended to stand as a comprehensive empirical study but to demonstrate usefulness of the G-StMAR model.

\section{Conclusions}\label{sec:conclusions}
This article introduced a mixture autoregressive model which is a combination of the Gaussian mixture autoregressive (GMAR) model \citep{Kalliovirta+Meitz+Saikkonen:2015} and the Student's $t$ mixture autoregressive (StMAR) model \citep{Meitz+Preve+Saikkonen:2018}. This model, referred to as the G-StMAR model, has several attractive theoretical and practical properties that are analogous to those of the GMAR and StMAR model. In addition to discussing the properties, it was noted that estimating the parameters of the G-StMAR model can be challenging in practice. Following \cite{Dorsey+Mayer:1995} \citep[and][]{Meitz+Preve+Saikkonen:2018,Meitz+Preve+Saikkonen2:2018}, we suggested using a two-phase estimation procedure where a genetic algorithm is used to find starting values for a gradient based method and accompied the paper with the R package uGMAR \citep{uGMAR} which implements the two-phase estimation procedure with a modified version of a genetic algorithm.

We stated that the G-StMAR model is a limiting case of a StMAR model with some degrees of freedom parameters tending to infinity, and found that large degrees of freedom estimates in a StMAR model are not only redundant but also cause several inconveniences in numerical analysis of the model. In particular, weak identification of large degrees of freedom parameters was found to lead to numerically nearly singular approximation of the observed information matrix when evaluated at the estimate, making the approximate standard errors for the estimates and \citeauthor{Kalliovirta:2012}'s (\citeyear{Kalliovirta:2012}) diagnostic tests often unavailable. Removing the redundant degrees of freedom parameters by switching to a G-StMAR model was concluded to obviate the problems. 

As an empirical application, we considered the monthly U.S. interest rate spread between the 3-month Treasury bill rate and the effective federal funds rate. Our G-StMAR model identified three regimes for the spread, with a switch from a StMAR type regime to a GMAR type regime arising from a switch in the economic regime, namely, to a regime where the zero lower bound limits the movements of the interest rates. The two StMAR type regimes accommodate eras of low mean and high variability and high mean and moderate variability. The first StMAR type regime arguably occurs often when the market anticipates decreases in the FF rate or possibly has increased preferences for safety, whereas the second one mostly occurs when the Fed is arguably not expected to significantly decrease the FF rate target. As opposed to modelling the series with a StMAR model containing an overly large degrees of freedom estimate, switching to the more parsimonious G-StMAR model allowed us to numerically compute approximate standard errors for the estimates, and moreover, to perform the \citeauthor{Kalliovirta:2012}'s (\citeyear{Kalliovirta:2012}) quantile residual tests which turned out to have significance in the model selection. 

\section*{Acknowledgements}
The author thanks Markku Lanne, Mika Meitz, and Pentti Saikkonen who commented the work and gave insightful suggestions that helped to improve the paper substantially. The author also thanks the Academy of Finland for financing the project.% (grant 308628).

\bibliography{refs.bib}

\begin{appendices}

\section{Modified genetic algorithm}
As discussed in Section \ref{sec:twophase}, the accompanied R package uGMAR \citep{uGMAR} employs a two-phase producedure for estimating the parameters of the G-StMAR model (and also of the GMAR \citep{Kalliovirta+Meitz+Saikkonen:2015} and the StMAR \citep{Meitz+Preve+Saikkonen:2018} model). In the first phase, a genetic algorithm is used to find starting values for a gradient based variable metric algorithm \citep[algorithm 21]{Nash:1990} which then, in the second phase, accurately converges to a nearby local maximum or saddle point. In this appendix, it is first briefly described how our version of the genetic algorithm functions in general, and then the specific modifications made to enhance estimation of the G-StMAR model are discussed \citep[for more detailed description of the genetic algorithm, see, e.g.,][]{Dorsey+Mayer:1995}.

In a genetic algorithm, an initial \textit{population} that consists of different parameter vectors (that are often drawn at random) is first constructed. Then the genetic algorithm operates iteratively so that in each iteration, referred to as \textit{generation}, the current population consisting of candidate solutions goes through the phases of \textit{selection}, \textit{crossover}, and \textit{mutation}. In the selection phase, parameter vectors are sampled with replacement from the current population to the \textit{reproduction pool} according to probabilities that are based on their \textit{fitness}, that is, on the related log-likelihoods. In the crossover phase, some of the parameter vectors in the reproduction pool are crossed over with each other, with the probabilities of experiencing crossover given by the \textit{crossover rate}. Finally, some of the parameter vectors are mutated in the mutation phase, with the mutation probabilities given by the \textit{mutation rate}. In our version of the genetic algorithm, mutation means that the mutating parameter vector is fully replaced with another parameter vector that is drawn at random \citep[in][mutations are drawn for each scalar component of parameter vectors individually]{Dorsey+Mayer:1995}. The reproduction pool that has experienced crossovers and mutations is the new population, and the algorithm proceeds to the next generation, \textit{evolving} towards the global maximum one generation after another.

Because the G-StMAR model can be challenging to estimate even with a robust estimation algorithm such as the genetic algorithm, we have made modifications to improve its performance. In particular, a slightly modified version\footnote{We modified it to enforce a 40\% minimum crossover rate for all individuals in the population} of the individually adaptive crossover rate and mutation rate introduced by \cite{Patnaik+Srinivas:1994} is employed in order to force the subaverage solutions to disrupt while protecting the better ones. The fitness inheritance proposed by \cite*{Smith+Dike+Stegmann:1995} is deployed to shorten the estimation time by cutting down the number computationally costly evaluations of the log-likelihood function. In order to enhance thorough exploration of the parameter space, the algorithm proposed by \cite{Monahan:1984} is used in some random mutations to generate parameter vectors near the boundary of the stationarity region. In the case of a premature convergence, most of the population is mutated so that exploration of the parameter space continues. Moreover, after a large number generations have been run, for faster convergence the random mutations will be targeted to a neighbourhood of the best-so-far parameter vector; we call these \textit{smart mutations}.   %After running the genetic algorithm, the estimation is finalized with a variable metric algorithm \cite[algorithm 21, implemented by \citealp{Rnormal}]{Nash:1990} using a central difference approximation for the gradient of the log-likelihood function.

In addition to the modifications described above, we have made further adjustments to care for the special structure of the log-likelihood function. Specifically, the definition of the mixing weights (\ref{gstmar:alphamt}) implies that if a regime has parameter values that fit poorly relative to the other regimes, the mixing weights drop to near zero. The surface of the log-likelihood function thus flattens in the related directions, meaning that the algorithm is unable to converge properly if the proposed parameter vectors don't pose a reasonable fit for all regimes. This problem of unidentified (or redundant) regimes often occurs when the number of mixture components is chosen too large, but it can be present even when the number of mixture components is chosen correctly. In uGMAR, we try to resolve this problem by penalizing parameter vectors containing redundant regimes with smaller probabilities to get chosen to the reproduction pool. Moreover, smart mutations are targeted only to the neighbourhood of parameter values that identify all regimes. If such parameter vectors have not been found (after a large number of generations have been run), combining regimes from different parameter vectors is attempted along with random search.

\section{Properties of multivariate Gaussian and Student's $t$-distribution}\label{sec:propertiesNorStu}
Denote a $d$-dimensional real valued vector by $\boldsymbol{y}$. It's well known that the density function of the $d$-dimensional multivariate Gaussian distribution with mean $\boldsymbol{\mu}$ and covariance matrix $\boldsymbol{\Gamma}$ is
\begin{equation}
n_d(\boldsymbol{y};\boldsymbol{\mu},\boldsymbol{\Gamma})=(2\pi)^{-d/2}\det(\boldsymbol{\Gamma})^{-1/2}\exp\left\lbrace -\frac{1}{2}(\boldsymbol{y} - \boldsymbol{\mu})'\boldsymbol{\Gamma}^{-1}(\boldsymbol{y} - \boldsymbol{\mu}) \right\rbrace.
\end{equation}

Similarly to \cite{Meitz+Preve+Saikkonen:2018} but differing from the standard form, we parametrize the Student's $t$-distribution using its covariance matrix as a parameter together with the mean and degrees of freedom. The density function of such a $d$-dimensional $t$-distribution with mean $\boldsymbol{\mu}$, covariance matrix $\boldsymbol{\Gamma}$, and $\nu>2$ degrees of freedom is
\begin{equation}
t_d\left(\boldsymbol{y};\boldsymbol{\mu},\boldsymbol{\Gamma},\nu\right)=C_d(\nu)\text{det}(\boldsymbol{\Gamma})^{-1/2}\left(1+\frac{(\boldsymbol{y}-\boldsymbol{\mu})'\boldsymbol{\Gamma}^{-1}(\boldsymbol{y}-\boldsymbol{\mu})}{\nu-2}\right)^{-(d+\nu)/2},
\end{equation}
where
\begin{equation}
C_d(\nu)=\frac{\Gamma\left(\frac{d+\nu}{2}\right)}{\sqrt{\pi^d(\nu-2)^d}\Gamma\left(\frac{\nu}{2}\right)},
\end{equation}
and $\Gamma\left(\cdot\right)$ is the gamma function. We assume that the covariance matrix $\boldsymbol{\Gamma}$ is positive definite for both distributions. 

Consider a partition $\boldsymbol{X}=(\boldsymbol{X}_1,\boldsymbol{X}_2)$ of either a normally or $t$-distributed (with $\nu$ degrees of freedom) random vector $\boldsymbol{X}$ such that $\boldsymbol{X}_1$ has dimension $(d_1\times1)$ and $\boldsymbol{X}_2$ has dimension $(d_2\times1)$. Consider also a corresponding partition of the mean vector $\boldsymbol{\mu}=(\boldsymbol{\mu}_1,\boldsymbol{\mu}_2)$ and the covariance matrix
\begin{equation}
\boldsymbol{\Gamma}=
\begin{bmatrix}
\boldsymbol{\Gamma}_{11} & \boldsymbol{\Gamma}_{12} \\
\boldsymbol{\Gamma}_{12}' & \boldsymbol{\Gamma}_{22}
\end{bmatrix},
\end{equation}
where, for example, the dimension of $\boldsymbol{\Gamma}_{11}$ is $(d_1\times d_1)$. Then in the case of normally distributed $\boldsymbol{X}$, $\boldsymbol{X}_1$ has the marginal distribution $n_{d_1}(\boldsymbol{\mu}_1,\boldsymbol{\Gamma}_{11})$ and $\boldsymbol{X}_2$ has the marginal distribution $n_{d_2}(\boldsymbol{\mu}_2,\boldsymbol{\Gamma}_{22})$. In the $t$-distributed case, the marginal distributions are $t_{d_1}(\boldsymbol{\mu}_1,\boldsymbol{\Gamma}_{11},\nu)$ and $t_{d_2}(\boldsymbol{\mu}_2,\boldsymbol{\Gamma}_{22},\nu)$ respectively (see, e.g., \cite{Ding:2016}, also in what follows).

In the normally distributed case, the conditional distribution of the random vector $\boldsymbol{X}_1$ given $\boldsymbol{X}_2=\boldsymbol{x}_2$ is 
\begin{equation}
\boldsymbol{X}_1\mid(\boldsymbol{X}_2=\boldsymbol{x}_2)\sim n_{d_1}(\boldsymbol{\mu}_{1\mid2}(\boldsymbol{x}_2),\boldsymbol{\Gamma}_{1\mid2}(\boldsymbol{x}_2))
\end{equation}
where
\begin{align}
\boldsymbol{\mu}_{1\mid2}(\boldsymbol{x}_2) &= \boldsymbol{\mu}_1+\boldsymbol{\Gamma}_{12}\boldsymbol{\Gamma}_{22}^{-1}(\boldsymbol{x}_2-\boldsymbol{\mu}_2) \quad \text{and} \\
\boldsymbol{\Gamma}_{1\mid2}(\boldsymbol{x}_2) &= \boldsymbol{\Gamma}_{11}-\boldsymbol{\Gamma}_{12}\boldsymbol{\Gamma}_{22}^{-1}\boldsymbol{\Gamma}_{12}'.
\end{align}
In the $t$-distributed case, the analogous conditional distribution is
\begin{equation}
\boldsymbol{X}_1\mid(\boldsymbol{X}_2=\boldsymbol{x}_2)\sim t_{d_1}(\boldsymbol{\mu}_{1\mid2}(\boldsymbol{x}_2),\boldsymbol{\Gamma}_{1\mid2}(\boldsymbol{x}_2),\nu+d_2),
\end{equation}
where
\begin{align}
\boldsymbol{\mu}_{1\mid2}(\boldsymbol{x}_2) &= \boldsymbol{\mu}_1+\boldsymbol{\Gamma}_{12}\boldsymbol{\Gamma}_{22}^{-1}(\boldsymbol{x}_2-\boldsymbol{\mu}_2) \quad \text{and} \nonumber\\
\boldsymbol{\Gamma}_{1\mid2}(\boldsymbol{x}_2) &= \frac{\nu-2+(\boldsymbol{x}_2-\boldsymbol{\mu}_2)'\boldsymbol{\Gamma}_{22}^{-1}(\boldsymbol{x}_2-\boldsymbol{\mu}_2)}{\nu-2+d_2}(\boldsymbol{\Gamma}_{11}-\boldsymbol{\Gamma}_{12}\boldsymbol{\Gamma}_{22}^{-1}\boldsymbol{\Gamma}_{12}'). \nonumber
\end{align}
In particular, we have
\begin{align}
n_d(\boldsymbol{x};\boldsymbol{\mu},\boldsymbol{\Gamma}) &=n_{d_1}(\boldsymbol{x}_1;\boldsymbol{\mu}_{1|2}(\boldsymbol{x}_2),\boldsymbol{\Gamma}_{1|2}(\boldsymbol{x}_2))n_{d_2}(\boldsymbol{x}_2;\boldsymbol{\mu}_2,\boldsymbol{\Gamma}_{22}) \quad \text{and}\\
t_d(\boldsymbol{x};\boldsymbol{\mu},\boldsymbol{\Gamma},\nu) &=t_{d_1}(\boldsymbol{x}_1;\boldsymbol{\mu}_{1|2}(\boldsymbol{x}_2),\boldsymbol{\Gamma}_{1|2}(\boldsymbol{x}_2),\nu+d_2)t_{d_2}(\boldsymbol{x}_2;\boldsymbol{\mu}_2,\boldsymbol{\Gamma}_{22},\nu).
\end{align}

\section{Proofs}

\subsection{Proof of Theorem 1}
Suppose $\lbrace y_t\rbrace_{t=1}^{\infty}$ is a G-StMAR process. Then the process $\boldsymbol{y}_t=(y_t,...,y_{t-p+1})$ is clearly a Markov chain on $\mathbb{R}^p$. Let $\boldsymbol{y}_0=(y_0,...,y_{-p+1})$ be a random vector whose distribution is  characterized by the density function $f(\boldsymbol{y}_0;\boldsymbol{\theta})=\sum_{m=1}^{M_1}\alpha_m n_p(\boldsymbol{y}_0;\mu_m\mathbf{1}_p,\boldsymbol{\Gamma}_{m,p})+\sum_{m=M_1+1}^{M}\alpha_m\times$ $t_p(\boldsymbol{y}_0;\mu_m\mathbf{1}_p,\boldsymbol{\Gamma}_{m,p},\nu_m)$. According to equations (\ref{eq:gausdistr1})-(\ref{eq:gausdistr3}), (\ref{eq:studentdist1})-(\ref{eq:studentdist3}), (\ref{gstmar:conddens}), and (\ref{gstmar:alphamt}), the density of the conditional distribution of $y_1$ given $\boldsymbol{y}_0$ is 
\begin{align}
\begin{aligned}
 f(y_1\mid\boldsymbol{y}_0;\boldsymbol{\theta}) & = \sum_{m=1}^{M_1}\frac{\alpha_m n_p(\boldsymbol{y}_0;\mu_m\mathbf{1}_p,\boldsymbol{\Gamma}_{m,p})}
{f(\boldsymbol{y}_0;\boldsymbol{\theta})}n_1(y_1;\mu_{m,1},\sigma^2_{m})\\
 &+ \sum_{m=M_1+1}^{M}\frac{\alpha_m t_p(\boldsymbol{y}_0;\mu_m\mathbf{1}_p,\boldsymbol{\Gamma}_{m,p},\nu_m)}
{f(\boldsymbol{y}_0;\boldsymbol{\theta})}t_1(y_1;\mu_{m,1},\sigma^2_{m,1},\nu_m+p)\\
\end{aligned}\\
\begin{aligned}
&= \sum_{m=1}^{M_1}\frac{\alpha_m}
{f(\boldsymbol{y}_0;\boldsymbol{\theta})}n_{p+1}((y_1,\boldsymbol{y}_0);\mu_m\mathbf{1}_{p+1},\boldsymbol{\Gamma}_{m,p+1}) \quad\quad\quad\quad\quad\quad\,\\
 &+ \sum_{m=M_1+1}^{M}\frac{\alpha_m}
{f(\boldsymbol{y}_0;\boldsymbol{\theta})}t_{p+1}((y_1,\boldsymbol{y}_0);\mu_m\mathbf{1}_{p+1},\boldsymbol{\Gamma}_{m,p+1},\nu_m).
\end{aligned}
\end{align}
The random vector $(y_1,\boldsymbol{y}_0)$ therefore has the density function
\begin{align}
\begin{aligned}
f((y_1,\boldsymbol{y}_0);\boldsymbol{\theta}) &=\sum_{m=1}^{M_1}\alpha_mn_{p+1}((y_1,\boldsymbol{y}_0);\mu_m\mathbf{1}_{p+1},\boldsymbol{\Gamma}_{m,p+1})\\
 &+\sum_{m=M_1+1}^{M}\alpha_m t_{p+1}((y_1,\boldsymbol{y}_0);\mu_m\mathbf{1}_{p+1},\boldsymbol{\Gamma}_{m,p+1},\nu_m).
\end{aligned}
\end{align}
Using properties of marginal densities of multivariate normal and $t$-distributions, by integrating $y_{-p+1}$ out we obtain the density of $\boldsymbol{y}_1$ as $f(\boldsymbol{y}_1;\boldsymbol{\theta})=\sum_{m=1}^{M_1}\alpha_m n_p(\boldsymbol{y}_1;\mu_m\mathbf{1}_p,\boldsymbol{\Gamma}_{m,p})+\sum_{m=M_1+1}^{M}\alpha_m\times$ $t_p(\boldsymbol{y}_1;\mu_m\mathbf{1}_p,\boldsymbol{\Gamma}_{m,p},\nu_m)$.\footnote{Because the covariance matrices $\boldsymbol{\Gamma}_{m,p+1}$ ($m=1,...,M$) have the Toepliz form and $\mu_m\mathbf{1}_p=(\mu_m,...,\mu_m)$, the marginal densities for random vectors shorter than $p$ are obtained by integrating the desired random variables out, and their distributions are mixtures of normal and $t$-distributions.} %Jos ei ole Toepliz muotoa, niin kovarianssi riippuu muustakin kuin etäisyydestä. -> y_0:n ja y_1:n samoinjakautuneisuus ei päde välttämättä. 
Thus, the random vectors $\boldsymbol{y}_0$ and $\boldsymbol{y}_1$ are identically distributed. As the process $\lbrace \boldsymbol{y}_t\rbrace_{t=1}^{\infty}$ is a (time homogeneous) Markov chain, it follows that $\lbrace \boldsymbol{y}_t\rbrace_{t=1}^{\infty}$ has a stationary distribution $\pi_{\boldsymbol{y}}(\cdot)$ characterized by the density $f(\cdot;\boldsymbol{\theta})=\sum_{m=1}^{M_1}\alpha_m n_p(\cdot;\mu_m\mathbf{1}_p,\boldsymbol{\Gamma}_{m,p})+\sum_{m=M_1+1}^{M}\alpha_m t_p(\cdot;\mu_m\mathbf{1}_p,\boldsymbol{\Gamma}_{m,p},\nu_m)$ \citep[pp. 230-231]{Meyn+Tweedie:2009}.

For ergodicity, let $P_{\boldsymbol{y}}^p(\boldsymbol{y},\cdot)=\mathbb{P}(\boldsymbol{y}_p\in\cdot|\boldsymbol{y}_0=\boldsymbol{y})$ signify the $p$-step transition probability measure of the process $\boldsymbol{y}_t$. Using the $p$th order Markov property of $y_t$, it's easy to check that $P_{\boldsymbol{y}}^p(\boldsymbol{y},\cdot)$ has the density 
\begin{equation}
f(\boldsymbol{y}_p|\boldsymbol{y}_0;\boldsymbol{\theta}) = \prod_{t=1}^{p}\left(\sum_{m=1}^{M_1}\alpha_{m,t} n_1(y_t;\mu_{m,t},\sigma^2_{m})+\sum_{m=M_1+1}^{M}\alpha_{m,t} t_1(y_t;\mu_{m,t},\sigma^2_{m,t},\nu_m+p) \right).
\end{equation}
Clearly $f(\boldsymbol{y}_p|\boldsymbol{y}_0;\boldsymbol{\theta})>0$ for all $\boldsymbol{y}_p\in\mathbb{R}^p$ and all $\boldsymbol{y}_0\in\mathbb{R}^p$, so we can conclude that $\boldsymbol{y}_t$ is ergodic in the sense of \citet[Ch. 13]{Meyn+Tweedie:2009} by using arguments identical to those used in the proof of Theorem 1 in \cite{Kalliovirta+Meitz+Saikkonen:2015}. $\blacksquare$

\subsection{Proof of Theorem 2}
First note that $L_T^{(c)}(\boldsymbol{\theta})$ is continuous, and that together with Assumption 1 of the main paper it implies existence of a measurable maximizer $\hat{\boldsymbol{\theta}}_T$. In order to conclude strong consistency of $\hat{\boldsymbol{\theta}}_T$, it needs to be shown that  \citep[see, e.g.,][Theorem 2.1 and the discussion on page 2122]{Newey+McFadden:1994}
\begin{enumerate}[label=(\roman*)]
\item the uniform strong law of large numbers holds for the log-likelihood function; that is,\\
 $\sup\limits_{\boldsymbol{\theta}\in\boldsymbol{\Theta}}\left\lvert L_T^{(c)}(\boldsymbol{\theta})-\text{E}\left[L_T^{(c)}(\boldsymbol{\theta})\right]\right\lvert\rightarrow 0$ almost surely as  $T\rightarrow\infty$,
\item and that the limit of $L_T^{(c)}(\boldsymbol{\theta})$ is uniquely maximized at $\boldsymbol{\theta}=\boldsymbol{\theta}_0$.
\end{enumerate}

\textbf{Proof of (i).}
Because the initial values are assumed to be from the stationary distribution, the process $\boldsymbol{y}_t=(y_t,...,y_{t-p+1})$, and hence also $y_t$, is stationary and ergodic, and $\text{E}\left[L_T^{(c)}(\boldsymbol{\theta})\right]=\text{E}\left[l_t(\boldsymbol{\theta})\right]$. To conclude (i), it thus suffices to show that $\text{E}\left[\sup_{\boldsymbol{\theta}\in\boldsymbol{\Theta}}\left\lvert l_t(\boldsymbol{\theta}) \right\lvert \right] < \infty$ \citep[see][]{Rao:1962}. This is done by using compactness of the parameter space to derive finite lower and upper bounds for $l_t(\boldsymbol{\theta})$ which is given by 
\begin{equation}\label{cons:lt}
l_t(\boldsymbol{\theta}) =\text{log}\left(\sum_{m=1}^{M_1}\alpha_{m,t}n_1(y_t;\mu_{m,t},\sigma_m^2)+\sum_{m=M_1+1}^{M}\alpha_{m,t} t_1\left(y_t;\mu_{m,t},\sigma_{m,t}^2,\nu_m+p\right)\right).
\end{equation}
We know from the structure of the parameter space that $c_1\leq\sigma_m^2\leq c_2$ and $c_1\leq\alpha_m\leq 1-c_1$ for all $m=1,...,M$, and $c_3\leq\nu_m\leq c_2$ for all  $m=M_1+1,...,M$, for some $0<c_1<1$, $c_2<\infty$ and $c_3>2$. Because the exponential function is bounded from above by one on the non-positive real axis, and in addition $c_1\leq\sigma_m^2$, there exists a constant $U_1<\infty$ such that 
\begin{equation}\label{cons:n1upper}
n_1(y_t;\mu_{m,t},\sigma_m^2)= \left(2\pi\sigma_{m}^2\right)^{-1/2}\text{exp}\left(-\frac{(y_t-\mu_{m,t})^2}{2\sigma_m^2}\right) \leq U_1
\end{equation}
for all $m=1,...,M_1$.

We also have $c_3 \leq\nu_m+p \leq c_2+p$ for all $m=M_1+1,...,M$. Combined with the fact that the Gamma function is continuous on the positive real axis, this implies that there exist constants $c_4>0$ and $c_5<\infty$ such that
\begin{equation}\label{cons:Cbounds}
c_4 \leq C_1(\nu_m+p)=\frac{\Gamma\left(\frac{1+\nu_m+p}{2}\right)}{\sqrt{\pi(\nu_m+p-2)}\Gamma\left(\frac{\nu_m+p}{2}\right)} \leq c_5
\end{equation}
for all $m=M_1+1,...,M$. Because $\boldsymbol{\Gamma}_m$ and hence $\boldsymbol{\Gamma}_m^{-1}$ is positive definite, $\sigma_m^2 \geq c_1$ and $c_3\leq\nu_m\leq c_2$, we can find some $c_6>0$ such that 
\begin{equation}\label{cons:sigmamt_lower}
\sigma_{m,t}^2=\frac{\nu_m-2+(\boldsymbol{y}_{t-1}-\mu_m\mathbf{1}_p)'\boldsymbol{\Gamma}_m^{-1}(\boldsymbol{y}_{t-1}-\mu_m\mathbf{1}_p)}{\nu_m-2+p}\sigma_m^2 \geq c_6
\end{equation}
for all $m=M_1+1,...,M$. Combined with (\ref{cons:Cbounds}) and (\ref{cons:sigmamt_lower}), the inequality $-(1+\nu_m+p)/2<0$ implies that there exists a constant $U_2<\infty$ for which
\begin{equation}\label{cons:t1upper}
t_1\left(y_t;\mu_{m,t},\sigma_{m,t}^2,\nu_m+p\right) =  \frac{C_1(\nu_m+p)}{\sigma_{m,t}} \left(1+\frac{(y_t-\mu_{m,t})^2}{(\nu_m+p-2)\sigma_{m,t}^2}\right)^{-(1+\nu_m+p)/2} \leq U_2.
\end{equation}
for all $m=M_1+1,...,M$. According to (\ref{cons:n1upper}), (\ref{cons:t1upper}) and the restriction $0\leq \alpha_{m,t} \leq 1$, there exists a constant $U_3<\infty$ such that 
\begin{equation}\label{cons:ltupper}
l_t(\boldsymbol{\theta})=\text{log}\left(\sum_{m=1}^{M_1}\alpha_{m,t}n_1(y_t;\mu_{m,t},\sigma_m^2)+\sum_{m=M_1+1}^{M}\alpha_{m,t} t_1\left(y_t;\mu_{m,t},\sigma_{m,t}^2,\nu_m+p\right)\right) \leq U_3.
\end{equation}

We know from compactness of the parameter space that
\begin{equation}
\frac{(y_t-\mu_{m,t})^2}{2\sigma_m^2} \leq c_7(1+y_t^2+\boldsymbol{y}_{t-1}'\boldsymbol{y}_{t-1}),
\end{equation}
implying
\begin{equation}
\exp\left\lbrace -\frac{(y_t-\mu_{m,t})^2}{2\sigma_m^2} \right\rbrace \geq \exp\left\lbrace -c_7(1+y_t^2+\boldsymbol{y}_{t-1}'\boldsymbol{y}_{t-1}) \right\rbrace,
\end{equation}
for all $m=1,...,M_1$, and for some finite constant $c_7$. By $\sigma_m^2 \leq c_2$ it also holds that $(2\pi\sigma_m^2)^{-1/2}\geq (2\pi c_2)^{-1/2}$, so
\begin{equation}\label{cons:n1lower}
n_1(y_t;\mu_{m,t},\sigma_m^2)\geq (2\pi c_2)^{-1/2}\exp\left\lbrace -c_7(1+y_t^2+\boldsymbol{y}_{t-1}'\boldsymbol{y}_{t-1}) \right\rbrace
\end{equation}
for all $m=1,...,M_1$.

Accordingly, since $\sigma_{m,t}^2 \geq c_6$ and $\nu_m \geq c_3$, it holds for some $c_8<\infty$ that 
\begin{equation}
1+\frac{(y_t-\mu_{m,t})^2}{(\nu_m+p-2)\sigma_{m,t}^2} \leq c_8(1+y_t^2+\boldsymbol{y}_{t-1}'\boldsymbol{y}_{t-1}), \enspace m=M_1+1,...,M.
\end{equation}
Thus, because $\nu_m \leq c_2$ and the inner functions below take values larger than one, we have
\begin{equation}\label{cons:t1lower1}
\left( 1+\frac{(y_t-\mu_{m,t})^2}{(\nu_m+p-2)\sigma_{m,t}^2} \right)^{-(1+\nu_m+p)/2}\geq \left( c_8(1+y_t^2+\boldsymbol{y}_{t-1}'\boldsymbol{y}_{t-1}) \right)^{-(1+c_2+p)/2}.
\end{equation}
As \cite{Meitz+Preve+Saikkonen:2018} state in the proof of Theorem 3, the quadratic form on the right-hand-side of (\ref{cons:sigmamt_lower}) satisfies 
\begin{equation}
(\boldsymbol{y}_{t-1}-\mu_m\mathbf{1}_p)'\boldsymbol{\Gamma}_m^{-1}(\boldsymbol{y}_{t-1}-\mu_m\mathbf{1}_p) \leq c_9(1+\boldsymbol{y}_{t-1}'\boldsymbol{y}_{t-1})
\end{equation}
for all $m=M_1+1,...,M$, and for some $c_9<\infty$. Since also $0< \nu_m-2 \leq c_2$ and  $\sigma_m^2 \leq c_2$, we have $\sigma_{m,t}^2 \leq c_{10}(1+\boldsymbol{y}_{t-1}'\boldsymbol{y}_{t-1})$ for some finite constant $c_{10}$. Combining the former inequality with (\ref{cons:Cbounds}) and (\ref{cons:t1lower1}) yields a lower bound
\begin{equation}\label{cons:t1lower}
t_1\left(y_t;\mu_{m,t},\sigma_{m,t}^2,\nu_m+p\right) \geq \frac{c_4}{(c_{10}(1+\boldsymbol{y}_{t-1}'\boldsymbol{y}_{t-1}))^{1/2}} \left( c_8(1+y_t^2+\boldsymbol{y}_{t-1}'\boldsymbol{y}_{t-1}) \right)^{-(1+c_2+p)/2}.
\end{equation}

Finally, the restriction $\sum_{m=1}^{M}\alpha_{m,t}=1$ together with (\ref{cons:n1lower}) and (\ref{cons:t1lower}) implies
\begin{align}\label{cons:ltlower}
\begin{aligned}
l_t(\boldsymbol{\theta}) &\geq \min \left\lbrace -\frac{1}{2}\log(2\pi)-\frac{1}{2}\log(c_2) -c_7(1+y_t^2+\boldsymbol{y}_{t-1}'\boldsymbol{y}_{t-1}), \right.\\
 &\left. \log(c_4)-\frac{1}{2}\log(c_{10}(1+y_t^2+\boldsymbol{y}_{t-1}'\boldsymbol{y}_{t-1})) - \frac{1+c_2+p}{2}\log\left(c_8(1+y_t^2+\boldsymbol{y}_{t-1}'\boldsymbol{y}_{t-1})\right) \right\rbrace .
\end{aligned}
\end{align}
As $\text{E}\left[y_t^2+\boldsymbol{y}_{t-1}'\boldsymbol{y}_{t-1}) \right] <\infty$ (because $y_t$ is stationary and has finite second moments), it follows from Jensen's inequality that 
\begin{equation}
\text{E}\left[ \log\left(c_8(1+y_t^2+\boldsymbol{y}_{t-1}'\boldsymbol{y}_{t-1})\right) \right] <\infty \text{ and } \text{E}\left[ \log\left(c_{10}(1+\boldsymbol{y}_{t-1}'\boldsymbol{y}_{t-1})\right) \right] <\infty.
\end{equation}
The upper bound (\ref{cons:ltupper}) together with (\ref{cons:ltlower}) and finiteness of the aforementioned expectations shows that $\text{E}\left[\sup_{(\boldsymbol{\theta},\boldsymbol{\nu})\in\boldsymbol{\Theta}}\left\lvert l_t(\boldsymbol{\theta}) \right\lvert \right]$ $< \infty$.  $\blacksquare$\\

\textbf{Proof of (ii).} Given that condition (\ref{gstmar:identcond}) of the main paper sets a unique order for the mixture components, proving that this identification condition is satisfied amounts to showing that $\text{E}\left[l_t(\boldsymbol{\theta})\right] \leq \text{E}\left[l_t(\boldsymbol{\theta}_0)\right]$, and that the equality $\text{E}\left[l_t(\boldsymbol{\theta})\right] = \text{E}\left[l_t(\boldsymbol{\theta}_0)\right]$ implies
\begin{align}
\begin{aligned}
\boldsymbol{\vartheta}_m=\boldsymbol{\vartheta}_{\tau_1(m),0} \text{ and } \alpha_m = \alpha_{\tau_1(m),0} \text{ when } m=1,...,M_1, \text{ and }\\
(\boldsymbol{\vartheta}_m,\nu_m)=(\boldsymbol{\vartheta}_{\tau_2(m),0},\nu_{\tau_2(m),0}) \text{ and } \alpha_m = \alpha_{\tau_2(m),0} \text{ when } m=M_1+1,...,M,
\end{aligned}
\end{align}
for some permutations $\lbrace \tau_1(1),...,\tau_1(M_1) \rbrace$ and $\lbrace \tau_2(M_1+1),...,\tau_2(M) \rbrace$. For notational clarity, we omit the subscripts from $y_t$ and $\boldsymbol{y}_{t-1}$, and write $\mu_{m,t}=\mu(\boldsymbol{y};\boldsymbol{\vartheta}_m)$, $\sigma_m^2=\sigma_m^2(\boldsymbol{\vartheta}_m)$, $\sigma_{m,t}^2=\sigma_{m,t}^2(\boldsymbol{y};\boldsymbol{\vartheta}_m,\nu_m)$ for the expressions in (\ref{cons:lt}) making clear their dependence on the parameter value. We leave the dependence of $\alpha_{m,t}$ on $\boldsymbol{\theta}$ and $\boldsymbol{y}$ unmarked and denote by $\alpha_{m,0,t}$ mixing weights based on the true parameter value.

Making use of the fact that the density function of $(y_t,\boldsymbol{y}_{t-1})$ has the form $f((y_t,\boldsymbol{y}_{t-1});\boldsymbol{\theta}) =\sum_{m=1}^{M_1}\alpha_m n_{p+1}((y_t,\boldsymbol{y}_{t-1}));\mu_m\mathbf{1}_{p+1},\boldsymbol{\Gamma}_{m,p+1})$
$+\sum_{m=M_1+1}^{M}\alpha_m t_{p+1}((y_t,\boldsymbol{y}_{t-1}));\mu_m\mathbf{1}_{p+1},\boldsymbol{\Gamma}_{m,p+1},\nu_m)$ (see proof of Theorem \ref{thm:stationarydist}) and reasoning based on Kullback-Leibler divergence, one can use arguments analogous to those in \citet[p. 265]{Kalliovirta+Meitz+Saikkonen:2015} to conclude $\text{E}\left[l_t(\boldsymbol{\theta})\right] - \text{E}\left[l_t(\boldsymbol{\theta}_0)\right] \leq 0$ with equality if and only if for almost all $(y,\boldsymbol{y})\in\mathbb{R}^{p+1}$
\begin{align}\label{ident:sames}
\begin{aligned}
&\sum_{m=1}^{M_1}\alpha_{m,t} n_1(y;\mu(\boldsymbol{y};\boldsymbol{\vartheta}_{m}),\sigma_m^2(\boldsymbol{\vartheta}_{m})) + \sum_{m=M_1+1}^{M}\alpha_{m,t} t_1(y;\mu(\boldsymbol{y};\boldsymbol{\vartheta}_m),\sigma_{m,t}^2(\boldsymbol{y};\boldsymbol{\vartheta}_{m},\nu_m)),\nu_{m}+p)\\
&=\sum_{m=1}^{M_1}\alpha_{m,0,t} n_1(y;\mu(\boldsymbol{y};\boldsymbol{\vartheta}_{m,0}),\sigma_m^2(\boldsymbol{\vartheta}_{m,0})) \\
&+ \sum_{m=M_1+1}^{M}\alpha_{m,0,t} t_1(y;\mu(\boldsymbol{y};\boldsymbol{\vartheta}_{m,0}),\sigma_{m,t}^2(\boldsymbol{y};\boldsymbol{\vartheta}_{m,0},\nu_{m,0})),\nu_{m,0}+p).
\end{aligned}
\end{align}
For each fixed $\boldsymbol{y}$ at a time, the mixing weights, conditional means and variances in (\ref{ident:sames}) are constants, so we may apply the result on identification of finite mixtures of normal and $t$-distributions in \citet[Example 1]{Holzmann+Munk+Gneiting:2006} (their parametrization of the $t$-distribution slightly differs from ours, but identification with their parametrization implies identification with our parametrization). For each fixed $\boldsymbol{y}$, there thus exists a permutation $\lbrace \tau_1(1),...,\tau_1(M_1) \rbrace$ (that may depend on $\boldsymbol{y}$) of the index set $\lbrace 1,...,M_1 \rbrace$ such that 
\begin{equation}\label{ident:gmar}
\alpha_{m,t}=\alpha_{\tau_1(m),0,t},\enspace \mu(\boldsymbol{y};\boldsymbol{\vartheta}_{m})=\mu(\boldsymbol{y};\boldsymbol{\vartheta}_{\tau_1(m),0}) \text{ and } \sigma_m^2(\boldsymbol{\vartheta}_{m})=\sigma_m^2(\boldsymbol{\vartheta}_{\tau_1(m),0})
\end{equation}
for almost all $y\in\mathbb{R}$ ($m=1,...,M_1$). Analogously, for each fixed $\boldsymbol{y}$ there exists a permutation $\lbrace \tau_2(M_1+1),...,\tau_2(M) \rbrace$ (that may depend on $\boldsymbol{y}$) of the index set $\lbrace M_1+1,...,M \rbrace$ such that 
\begin{align}\label{ident:stmar}
\begin{aligned}
&\nu_m = \nu_{\tau_2(m),0}, \enspace \alpha_{m,t}=\alpha_{\tau_2(m),0,t},\enspace \mu(\boldsymbol{y};\boldsymbol{\vartheta_{m}})=\mu(\boldsymbol{y};\boldsymbol{\vartheta}_{\tau_2(m),0})\\
&\text{ and } \sigma_{m,t}^2(\boldsymbol{y};\boldsymbol{\vartheta}_{m},\nu_m)=\sigma_{m,t}^2(\boldsymbol{y};\boldsymbol{\vartheta}_{\tau_2(m),0},\nu_{\tau_2(m),0}),
\end{aligned}
\end{align}
for almost all $y\in\mathbb{R}$ ($m=M_1+1,...,M$).

As argued by \citet[pp. 265-266]{Kalliovirta+Meitz+Saikkonen:2015}  for the GMAR type components, it follows from (\ref{ident:gmar}) that $\boldsymbol{\vartheta}_m=\boldsymbol{\vartheta}_{\tau_1(m),0}$ and $\alpha_m=\alpha_{\tau_1(m),0}$ for $m=1,...,M_1$. Accordingly, \cite{Meitz+Preve+Saikkonen:2018} showed that (\ref{ident:stmar}) implies $\boldsymbol{\vartheta}_m=\boldsymbol{\vartheta}_{\tau_2(m),0}$, $\nu_m = \nu_{\tau_2(m),0}$ and $\alpha_m=\alpha_{\tau_2(m),0}$ for $m=M_1+1,...,M$, completing the proof of strong consistency. 

Given consistency and assumptions of the theorem, asymptotic normality of the ML estimator can now be concluded using standard arguments. The required steps can be found, for example, in \citet*[proof of Theorem 3]{Kalliovirta+Meitz+Saikkonen:2016}. We omit the details for brevity. $\blacksquare$

\end{appendices}

\end{document}